\documentclass{article}

\usepackage{PRIMEarxiv}

\usepackage[utf8]{inputenc} 
\usepackage[T1]{fontenc}    
\usepackage{hyperref}       
\usepackage{url}            
\usepackage{booktabs}       
\usepackage{amsfonts}       
\usepackage{nicefrac}       
\usepackage{microtype}      
\usepackage{lipsum}
\usepackage{fancyhdr}       
\usepackage{graphicx}       
\graphicspath{{media/}}     
\usepackage{CJKutf8}
\usepackage{makecell}
\usepackage{multirow}
\usepackage{array}
\usepackage{booktabs}
\usepackage{hhline}
\usepackage{amsmath}
\usepackage{caption}
\usepackage{capt-of}

\title{AttenTrack: Mobile User Attention Awareness Based on Context and External Distractions
\thanks{\textit{\underline{Citation}}: 
\textbf{Authors. Title. Pages.... DOI:000000/11111.}} 
}

\author{
  Yutong Lin, Suyuan Liu \\
  University of Science and Technology of China \\
  \texttt{\{linyutong, lsysue\}@mail.ustc.edu.cn} \\
   \And
  Kaiwen Guo \\
  Ocean University of China \\
  \texttt{kevinguo@ouc.edu.cn} \\
     \And
  Haohua Du \\
  Beihang University \\
  \texttt{duhaohua@buaa.edu.cn} \\
     \And
  Chao Liu \\
  Ocean University of China \\
  \texttt{liuchao@ouc.edu.cn} \\
     \And
  Xiang-Yang Li \\
  University of Science and Technology of China \\
  \texttt{Xiangyang Li@ustc.edu.cn} \\
}
\begin{document}
\maketitle

\begin{abstract}
In the mobile internet era, managing limited attention amid information overload is crucial for enhancing collaboration and information delivery. However, current attention-aware systems often depend on wearables or personalized data, limiting their scalability and cross-context adaptability.

Inspired by psychological theories, we attempt to treat mobile notifications as naturally occurring external distractions and infer users’ attention states based on their response behaviors and contextual information. 
Our goal is to build an attention-aware model that does not rely on personalized historical data or complex subjective input, while ensuring strong cold-start capability and cross-context adaptability.

To this end, We design a field study framework integrating subjective and objective data, closely aligned with real-world external distractions (i.e., mobile notifications). 
Through field studies, we construct a fine-grained and interpretable dataset centered on the relationship among \textit{current context – external distractions – subjective attention}. 
Through our field studies, we conduct an in-depth analysis of the relationships among users’ response behaviors, response motivations, contextual information, and attention states. 
Building on our findings, we propose AttenTrack, a lightweight, privacy-friendly attention awareness model with strong cold-start capability. 
The model relies solely on non-privacy-sensitive objective data available on mobile devices, and can be applied to a variety of attention management tasks. 
In addition, we will publicly release the constructed dataset to support future research and advance the field of mobile attention awareness\footnote{\url{https://anonymous.4open.science/r/AttenTrack-6C0B/}}.
\end{abstract}


\section{Introduction}
\label{sec:introduction}
The user's \textbf{attention state} is typically defined as the degree of engagement in a given task at a specific moment. 
Given the limited nature of cognitive resources, attention is also regarded as a psychological regulation process that actively suppresses distractions and focuses on goal-relevant information~\cite{luck2002attention, styles2006psychology}. 
In an era where mobile computing and social media have profoundly transformed the information landscape, users’ limited attentional resources are constantly challenged by the overwhelming influx of competing information.
Consequently, the ability to sense and support users in managing their attention more effectively has emerged as a critical research focus — essential for enhancing collaborative efficiency in mobile environments, facilitating health-related interventions, and optimizing the overall information consumption experience.

In recent years, researchers have explored various attention-aware methods from multiple dimensions, covering interdisciplinary domains and complex scenarios~\cite{2010Oasis, park2017don, dabbish2004controlling, tanaka2021estimating, lindhiem2022objective, ahuja2019edusense}. 
However, most existing approaches still face significant challenges, such as complex deployment requirements and poor generalizability across users and contexts.

Research in non-mobile contexts often relies on controlled environments or explores physiological and behavioral representations of attention using multimodal data. 
For example, sensors such as Electroencephalography (EEG), Electrodermal Activity (EDA), and eye-tracking are widely used for cognitive state recognition~\cite{prabhu2023real, di2018unobtrusive, fan2021safedrivin}. 
However, these methods face limitations in practical applications due to the intrusiveness of the devices and the user burden, making them difficult to scale widely.

With the widespread adoption of smartphones, mobile contexts have become a new focus for research on individual cognitive resources. 
Existing studies have explored the impact of emotional states, social factors, and task contexts on user attention, leveraging perceptible information from smartphones to model attention. 
One line of research focuses on attention switching costs by detecting ``breakpoints'' or interruptions in user behavior and cognition, aiming to optimize notification strategies~\cite{forlivesi2018mindful, 2005Using, okoshi2015attelia}. 
However, methods centered on interrupting and redirecting attention are challenging to directly apply to systems designed to support sustained attention maintenance. 
Another line of research predicts users' level of engagement with specific content or applications based on their behavioral logs, but such approaches are mostly limited to digital media or social networking applications and lack cross-context, system-level scalability~\cite{mathur2016engagement, olaniyi2022user, huynh2018engagemon, tian2021and}. 
Moreover, some studies incorporate personalized information, such as personality traits, geographical location, and notification content, to enhance system intelligence~\cite{0How, 2014InterruptMe, 0Smartphone, 2015Designing}. 
However, the reliance on sensitive data raises privacy concerns and restricts the applicability of such systems in privacy-constrained environments, further weakening their ability to serve new or data-scarce users in cold-start situations.

Psychological research shows that attention fluctuations in users under similar contexts follow certain patterns, and users typically exhibit stronger resistance to distractions when highly engaged. 

\textbf{This leads us to hypothesize that} the resistance to distractions displayed by users in highly focused states can be observed in their interactions with mobile notifications. 
Consequently, we attempt to treat mobile notifications as naturally occurring external distractions, and by analyzing user responses to these distractions along with their current contextual information, we aim to uncover attention fluctuation patterns at the broader group level. 
This approach enables the development of an attention-aware system that does not rely on personalized historical data or complex subjective feedback, offering strong cold-start capabilities and cross-context adaptability. 
At the same time, we emphasize that interactions between users and external distractions, as well as attention states themselves, are multidimensional and complex, and should be explored from various perspectives.

In simple terms, we describe the user's attention state as a momentary level of focus, denoted as \( A \). 
Our goal is to train a model \( F \) such that, given contextual information  \( \boldsymbol{C} \in \mathbb{R}^m \) and external distraction-related information \( \boldsymbol{D} \in \mathbb{R}^n \), it can accurately predict the user's current attention states: \( \hat{A} = F(C, D; \theta) \), where \( \theta \) represents the model parameters.

However, uncovering the implicit connections among \textbf{\textit{current context – external distractions – subjective attention}}—and enabling attention awareness that generalizes across users and scenarios—remains a significant challenge:

First, as attention is a subjective and highly dynamic mental state, capturing it in natural settings requires minimizing experimental interference while ensuring both ecological validity and data reliability.

In addition, although a user's attention state influences their response to external distractions, this behavior is also affected by multiple factors such as social context, behavioral habits, and attributes of the distraction. 
The underlying mechanism of the response behavior can be abstracted as:
$d_{\text{response behavior}} = a f(A) + \beta g(S) + \varepsilon$,
where $S$ denotes the attributes of the distraction, and $\varepsilon$ represents unknown perturbing factors. Therefore, users' responses to distractions are driven by complex motivations and thus warrant further in-depth investigation.

Finally, the model must possess strong cold-start capability, maintaining high accuracy and stability even in the absence of personalized prior knowledge, and in the face of user variability and dynamic environmental conditions.

To address the aforementioned challenges, we develop a field study framework that combines both objective and subjective perspectives. 
By tightly coupling data collection with notification delivery, we achieve a strong correlation between data collection and real-world external distractions while minimizing additional experimental interference. 
In parallel with the automated collection of objective data, we adopt an open design to capture users' genuine motivational responses to notification interruptions. 
Through multiple rounds of field study to collect rich information, we create a highly interpretable and fine-grained dataset of \textbf{\textit{current context–external distractions–subjective attention}} tailored for mobile settings.

Based on field study, we systematically explore the intrinsic relationships between various contextual information and users' attention states, and further analyze the connections between the nature of external distractions, attention states, and users' response behaviors. 

Specifically, we employ a human-machine collaborative approach, utilizing large language models (LLM) to assist in the qualitative coding and quantitative analysis of the real response motivations collected during the field study.
The analysis reveals that approximately 55\% of users' interactions with external distractions (i.e., mobile notifications) are driven by personal state factors. 
Furthermore, we identified 29 behavioral patterns that influence user response behaviors. 
Through an in-depth analysis of the relationships between these 29 patterns and 6 types of fine-grained response behaviors, we uncovered the underlying mechanisms of attention states in responding to external distractions, enriching our understanding of the motivations behind user response behaviors. 
Additionally, we examined the role of distraction attributes (such as notification origin). 
The results show that user response behaviors are closely related to their attention states. 
However, in low-focus states, response behaviors are more strongly driven by the attributes of the distraction, particularly notifications with communication-related characteristics or higher content novelty, which tend to attract user responses due to their inherent features.

Based on the theoretical analysis results, we developed a lightweight, privacy-friendly, and cold-start capable focus prediction model, \textbf{AttenTrack}. 
The model is built solely on non-sensitive objective data from the smartphone, without the need for complex physiological signals or prior subjective user knowledge. 
By simply adjusting the threshold division method, AttenTrack can be applied to attention-maintenance scenarios, interruption-optimization scenarios, and fine-grained attention detection scenarios, demonstrating excellent scalability and cold-start capabilities.

Experiments show that in interruption-optimization scenarios, AttenTrack achieves an average F1 score of 80.09\% and a recall of 85.84\%, outperforming the baseline by over 23\%. 
In attention-maintenance and fine-grained attention detection scenarios, the model improves AUC by 20\% compared to the baseline. 
Personalized evaluation further reveals that building group models tailored to specific user populations or incorporating limited user-specific prior knowledge can further enhance AttenTrack’s performance.

We further examine how AttenTrack can be seamlessly adapted to various attention management tasks in mobile environments, such as group collaboration platforms and health management platforms. 
To support future research and advance the field, we will publicly release the dataset constructed in this study\footnote{\url{https://anonymous.4open.science/r/AttenTrack-6C0B/}}. 
To the best of our knowledge, this is the first publicly available dataset on user attention states collected via smartphones in real-world mobile settings.

\section{Related Work}
\label{sec:related_work}
Attention is a complex and multidimensional construct, commonly defined as the degree of a user’s engagement with the current task at a given moment or the selective processing of incoming information~\cite{farmer2019cognition, james1890principles}. 
Its core characteristic lies in the limitation of cognitive resources, and it is therefore also described as the ability to ignore irrelevant information~\cite{luck2002attention, styles2006psychology}.

Given the limited nature of attentional resources, prior studies have explored how to sense and model individuals' attention states from multiple dimensions, including contextual information, user behavioral features, and physiological signals, to support collaboration efficiency and enhance human-computer interaction~\cite{2010Oasis, horvitz2004busybody, 2004If, park2017don, dabbish2004controlling, tanaka2021estimating}. 
Below, we first review research progress in non-mobile settings, then turn to user attention studies in the context of mobile devices, and finally introduce the core focus of our work.

\subsection{Attention Research in Non-Mobile Contexts}
Attention research in non-mobile settings typically relies on relatively closed and controlled environments (e.g., cockpits, offices, classrooms) to explore representations of users' attention states through multimodal data~\cite{lindhiem2022objective, sankesarapassive, ahuja2019edusense}. 
For example, in office environments, users’ work engagement can be assessed by integrating PC interaction logs, online schedules, and audio-visual information captured by cameras and microphones~\cite{2004Learning, 2003Learning, horvitz2004busybody, iqbal2010notifications}.

In addition, psychophysiological measurement plays a significant role in this field. Tools such as EEG, EDA sensors, and eye-tracking devices are commonly used to explore causal links between physiological signals and psychological states~\cite{abdelrahman2019classifying}. 
For example, EEG signals are used to assist in ADHD diagnosis~\cite{prabhu2023real}, wearable EDA devices help assess students’ or audiences’ engagement with content~\cite{di2018unobtrusive, gao2020n}, and electromyography (EMG) signals are applied to detect abnormal behaviors during driving~\cite{fan2021safedrivin}. 
Although these physiological sensors provide high-precision and continuous monitoring, their wearable or intrusive nature limits their scalability and deployment in daily life.

\subsection{Attention Research in Mobile Contexts}
As mobile devices have become the primary medium for information access and social interaction, attention research in mobile contexts has gained increasing interest, particularly for advancing human-computer interaction and supporting group collaboration~\cite{cavdar2020multi, zulkernain2010mobile}. 
Current studies focus on understanding and modeling users’ attention states, exploring through controlled experiments and field studies how individuals’ attentional resources in mobile environments are influenced by psychological states, social factors, and task contexts~\cite{mehrotra2016my, pielot2017beyond}. 
These theoretical insights inform the development of more effective attention modeling approaches, ultimately enabling better attention management.

In the context of mobile notifications, attention-aware research typically focuses on the modeling and management of interruptibility~\cite{2017Understanding, tanaka2011study}. 
Interruptibility is defined from a system perspective as the appropriateness of interrupting the user at a given moment. 
Leveraging the rich sensors and behavioral logs available on smartphones, systems can infer users’ physical activities, environmental conditions, and phone usage patterns. 
Building on this, some studies draw from the psychological concept of breakpoints, proposing ``delayed until breakpoint'' strategies that aim to identify natural transition points between user tasks and activities, thereby optimizing information delivery and minimizing disruption~\cite{2004If, park2017don, forlivesi2018mindful, 2005Using, okoshi2015attelia}. 
Using the ESM, many works have further explored multiple factors influencing users’ perception and response to interruptions, including task complexity, environmental context, and the importance and interest level of the interrupting content, enriching the user-centric perspective in interruptibility modeling~\cite{2007Xensible, 2017Attention, 2015Designing}. 
However, these approaches often rely on guided questions or predefined options and are typically not tightly linked to actual interruption events, which limits users’ freedom to express their genuine interruption experiences~\cite{2014InterruptMe, 0How}. 
Moreover, existing interruptibility research is largely oriented towards facilitating attentional shifts, making it less applicable to systems that aim to support sustained attention maintenance.

In the domain of digital media and social networks, attention-aware research primarily centers on modeling user engagement, focusing on users’ attentional investment in specific content or applications~\cite{mathur2016engagement, olaniyi2022user}. 
For example, existing studies predict users’ immersion levels in applications based on online behavioral metrics, such as app usage logs, or detect users’ engagement in mobile gaming through touch event patterns captured by smartphones~\cite{huynh2018engagemon, tian2021and}. 
Such research typically targets specific contexts and tasks, making it suitable for attention management at the application level, but its applicability remains limited for system-level or cross-context attention management.

Meanwhile, some attention-aware studies incorporate more privacy-sensitive personalized features, such as users’ personality traits, location categories, relationships with notification senders, and notification content, to enhance the system’s understanding of users' psychological preferences and behavioral tendencies~\cite{0How, 2014InterruptMe, 0Smartphone, 2015Designing}. 
While the integration of such personalized features helps improve the intelligence of attention models, the heavy reliance on sensitive information may raise users' privacy concerns and limit the applicability of these systems in privacy-constrained contexts. 
More critically, such approaches that depend on personalized prior information often face cold-start challenges when dealing with new users or users with sparse data, leading to limited model adaptability.

Similar to existing research, we also explore the feasibility of modeling users’ subjective attention states in mobile contexts based on perceivable objective data. 
However, unlike prior work that focuses on ``interruptibility'' or ``user engagement'', we aim to develop a more generalizable attention prediction framework. 
Specifically, we focus on users’ \textbf{focused attention}, referring to the degree of concentration an individual dedicates to a specific task. Our aim is to capture the internal allocation of cognitive resources without limiting the study to particular contexts. 
We further aim to overcome the reliance on personalized information and prior user feedback seen in existing methods, instead mining population-level behavioral patterns to infer individual attention dynamics. 
This design enhances the model's adaptability in cold-start and privacy-constrained scenarios, enabling it to support not only interruption management but also applications that require sustained cognitive engagement.


\section{Study Methodology}
\label{sec:methodology}
We conducted two rounds of field studies, leveraging smartphones to deeply explore variations in users' attention across different times and contexts. Both rounds of studies are facilitated by a study app we developed in-house, which participants installed on their personal smartphones for remote data collection. 

We first conducted a pilot study, followed by two formal studies and we focus exclusively on formal studies in this work, analyzing and utilizing data from them.

\subsection{Core of Design} 
Extensive research has shown that EDA can effectively reflect emotional states and levels of attention~\cite{hernandez2014using, hernandez2013measuring, silveira2013predicting}. Some prior studies also utilize EDA to explore user engagement. 
However, in this study, we aim to investigate variations in users'  attention within unconstrained, real-world environments. 
To ensure that the data collection process integrates more seamlessly into participants' daily lives and improve generalizability, we choose not to rely on wearable devices.
Instead, we utilize smartphones to deliver ESM surveys, capturing users' attention at randomly sampled moments.

The core objective of our study application design is to enable continuous data collection while closely aligning with participants' daily routines. The key challenge lies in minimizing the psychological burden of sampling on participants to avoid disrupting the authenticity of their responses.

Rather than sending ESM surveys at pre-scheduled times, we align their delivery with moments when users receive notifications.
This design offers two key advantages. 
First, it avoids introducing new distractions at additional time points, minimizing experimental intrusion. Compared to ``forced interruptions'', it enhances the authenticity of users' responses. 
Secondly, it allows us to observe user interactions with notifications as external distractions, focusing on their responses, which helps us better understand users' attention dynamics in real-world distraction events.

\begin{figure}[h]
  \centering

\includegraphics[width=0.55\linewidth]{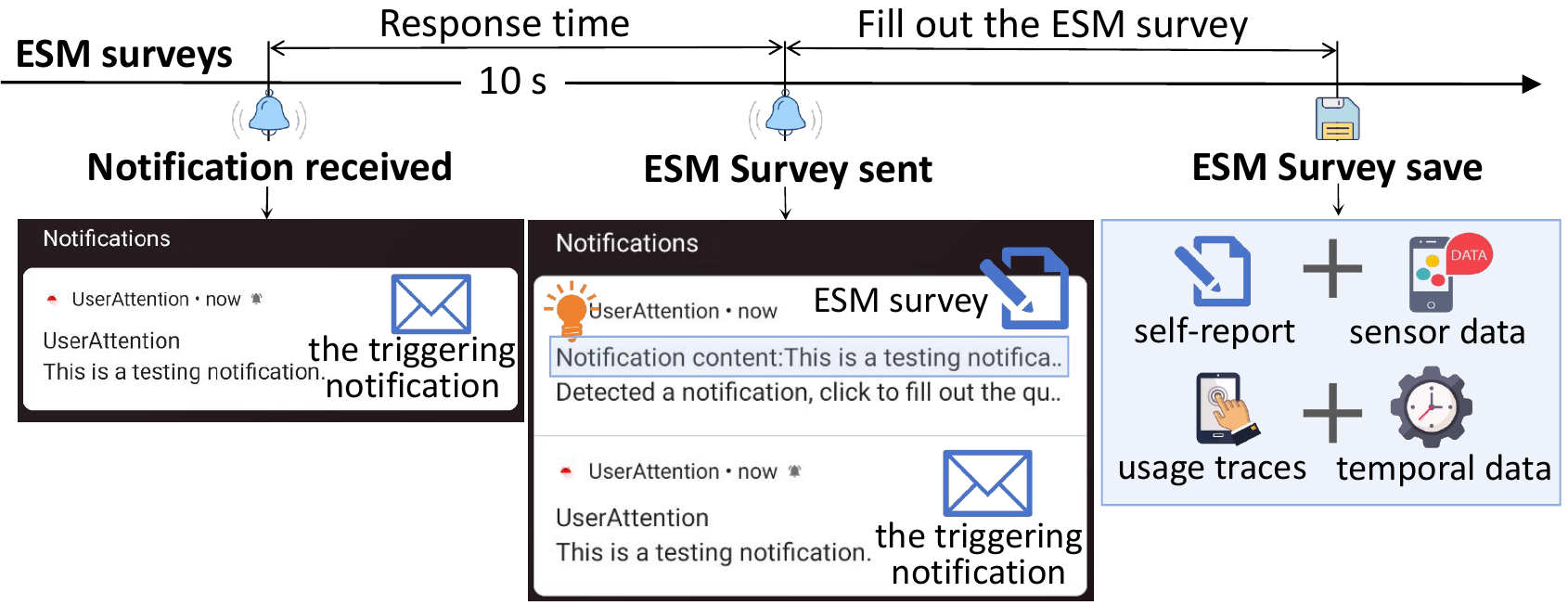}
  \caption{The figure depicts the timeline for data collection through ESM surveys. The ESM survey indicates the triggering notification in the title, helping users match it for response.}
      \label{fig1}
\end{figure}

\subsection{Study App}
We designed and developed an Android app for remote data collection. To ensure participants are well-informed about collected data and can use the app effectively, we developed a comprehensive user manual. 
Additionally, the app features a step-by-step onboarding process that helps participants grant the necessary permissions. 

During the study, participants were prompted to fill out ESM surveys through the app, which were triggered by notification events. The app consistently collects self-reported data from these surveys, in addition to various sensor data and phone usage logs.

The application runs in the background. Upon detecting a new notification, the app automatically triggers an ESM survey prompt for the participants. As illustrated in Figure~\ref{fig1}, there is a 10-second delay between the detection of a new notification and the dispatch of the survey notification. This delay ensures that the survey does not obscure the notification, allowing users to view their notifications without disruption and minimizing any potential influence on their behavior.
Once an ESM survey notification is sent, it remains in the notification tray. However, if multiple surveys are triggered in a short time frame, only the most recent one is kept. 

\subsubsection{Participant Privacy Controls}
\
\newline
To protect participant privacy, we implement several measures. 
The user manual explains the data collection process and required permissions. 
After each ESM survey, participants can review and decide whether to keep any automatically collected data.
All data is stored on participants' devices, with transmission to us occurring only upon study completion. 
Notably,the study does not involve collecting the notifications content.

\subsubsection{Collected Data}
\begin{figure}[ht]
  \centering
 \includegraphics[width=0.7\linewidth]{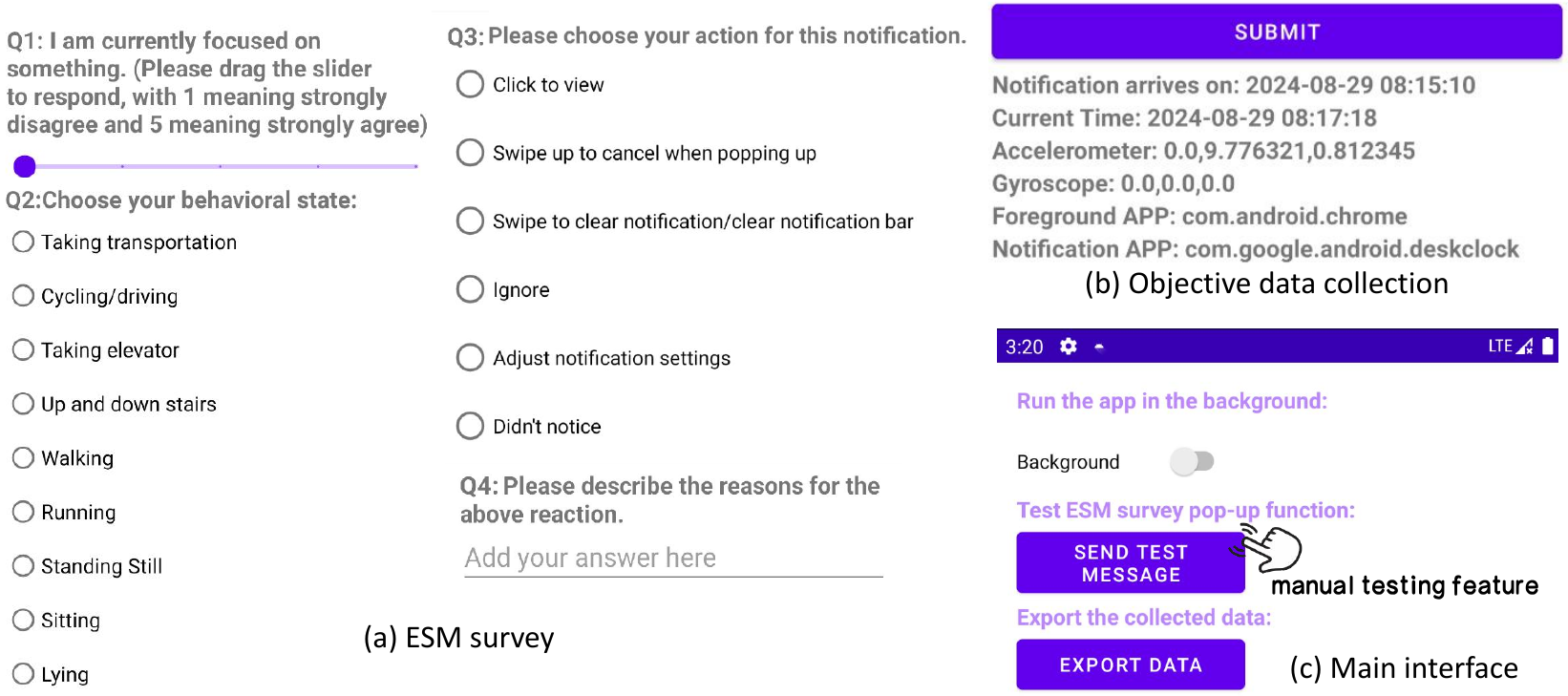}
  \caption{The figure illustrates the questionnaire content of the ESM survey, the automatically collected data, and the main interface of the study app.}
     \label{fig:ESM}
  \vspace{-6pt}
\end{figure}

\
\newline
ESM~\cite{hektner2007experience} collects self-reports on participants' behaviors and psychological states in daily life.
Figure~\ref{fig:ESM} shows a screenshot of our survey presented to participants.

Our ESM survey consists of four questions, with the key question being Q1 ``I am currently focused on something. (Please drag the slider to respond, with 1 meaning strongly disagree and 5 meaning strongly agree).'' We use a Likert scale to measure participants' attention, and they can select one of five levels by adjusting the slider, with higher values indicating greater focus.

Q2 inquires about participants' current physical activity. While all listed options are detectable via the HarmonyOS Motion API, self-reports are used to enhance accuracy and reliability.

Q3 and Q4 ask participants to indicate how they handled the notification that triggered the survey and briefly explain why.
Although Android's NotificationListenerService can automatically track interactions (e.g., ignored, tapped, swiped away, or cleared), we chose to collect self-reported responses. 
This design not only encourages participants to reflect on their motivations before answering Q4 but also provides more detailed handing behavioral data, allowing deeper analysis of user intent. Unlike studies using leading questions or predefined options, we used open-ended questions to enhance ecological validity and uncover new insights.

While continuous tracking offers richer data, it increases battery use and may burden users due to data invisibility. To balance quality and user experience, we collect objective data only during ESM submissions and display it afterward to ensure transparency.
All the objective data we collected are presented in Table~\ref{tab:Data_description}.

\subsection{Participants}
Participants were recruited online, targeting active, experienced Android users. 
In the first round of the formal study, 20 participants provided valid data after consenting to the study. 
After confirming the data’s effectiveness, we launched a second round to increase diversity, recruiting 23 participants. 
Details are provided in Appendix \ref{sec:appendix_model}. To maintain engagement, participants were compensated based on their valid data volume.

\subsection{Study Procedure}
We conducted a pilot study with six participants using diverse phones and their feedback on the app and design was collected and addressed before starting the formal study.

We standardized our field study into three consistent phases. 
First, participants were recruited, briefed on the study's purpose and procedures, and signed the informed consent form.
Second, they installed and calibrated the app, guided by detailed instructions, with three days for setup and trial runs. 
Finally, formal data collection began.

\subsection{Dataset}
\begin{figure}[ht]
  \centering
  \begin{minipage}[c]{0.76\textwidth}
        \centering
    \captionof{table}{Data types and descriptions. Bracketed values under ``attention states'' show data volume per state; notably, data decreases as attention rises.}
    \label{tab:Data_description}
  \small
  \begin{tabular}{|c|c|}
    \hline
    \textbf{Data Type} & \textbf{Description} \\
    \hline
    Temporal & 
    \begin{tabular}[c]{@{}l@{}}
        Time of day \{morning, afternoon, evening, night\}, response time, \\
        day of week, ESM \{receive, click\}, weekday \{true, false\}
    \end{tabular} \\
    \hline
    \begin{tabular}[c]{@{}c@{}}Physical\\activity\end{tabular} & 
    \begin{tabular}[c]{@{}l@{}}
        Taking transportation, cycling, driving, taking elevator, lying, \\
        up and down stairs, running, standing still, sitting, walking
    \end{tabular} \\
    \hline
    Sensors & Acceleration, gyroscope \\
    \hline
    Application & 
    Foreground/notification application name and category\\
    \hline
    \begin{tabular}[c]{@{}l@{}}Response\\behavior \end{tabular}&
    \begin{tabular}[c]{@{}l@{}}
        Click to view, swipe to clear notification/clear notification bar,  \\ 
        swipe up to cancel when popping up,adjust notification settings
    \end{tabular} \\
    \hline
    Motivation &
    \begin{tabular}[c]{@{}l@{}}
    Open-ended question: freely respond to the motivations behind the\\
    handing behavior to notifications.
    \end{tabular} \\
    \hline
    \begin{tabular}[c]{@{}c@{}}Attention\\state \end{tabular} & 
    \begin{tabular}[c]{@{}l@{}}
        1: highly unfocused (3366), 2: unfocused (1362), \\
        3: moderate (1586), 4: focused (1344), 5: highly focused (1350)
    \end{tabular} \\
    \hline
  \end{tabular}
  \end{minipage}
  \hfill 
  \begin{minipage}[c]{0.2\textwidth}
    \centering
      \includegraphics[width=\textwidth]{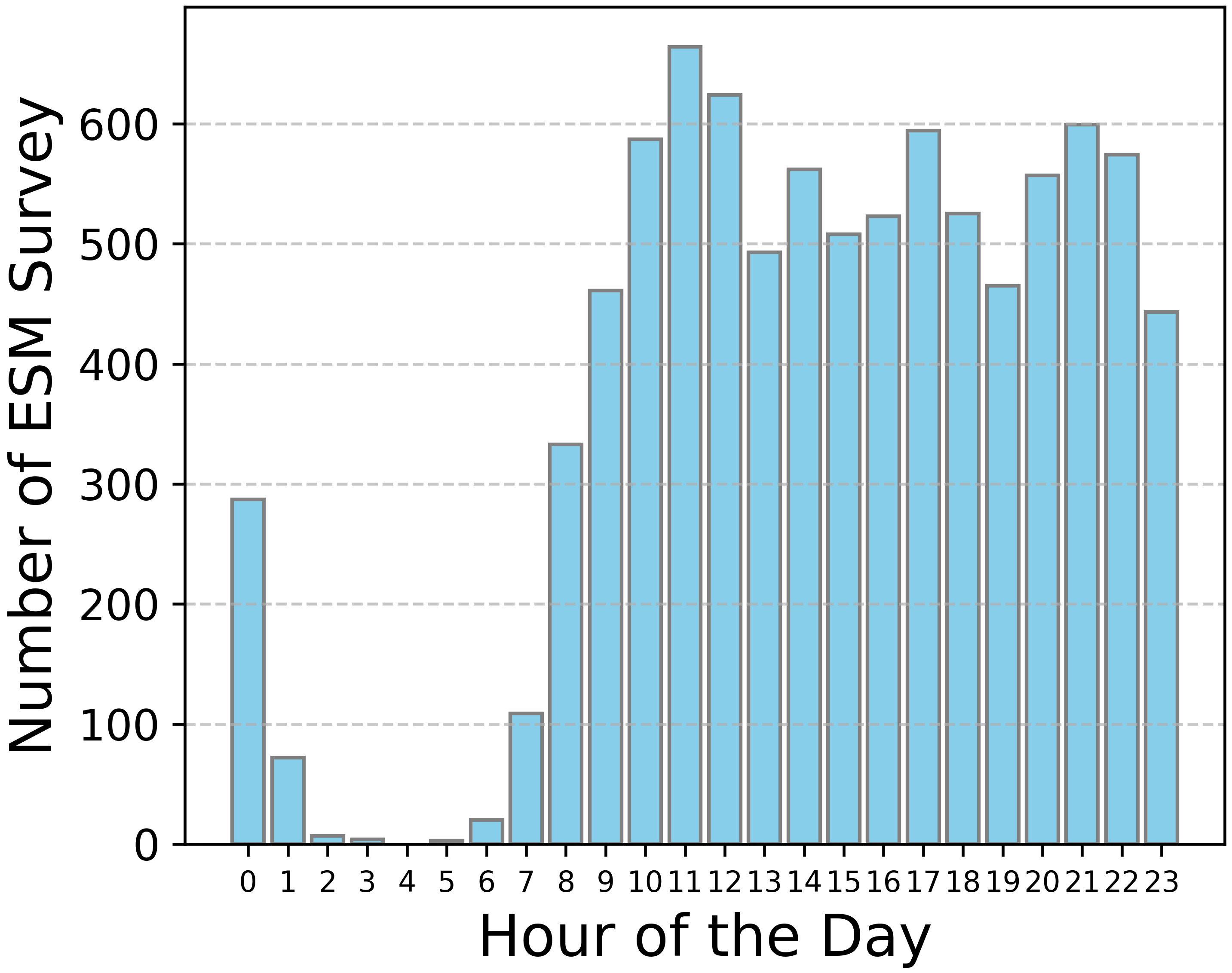}
      \captionof{figure}{The bar chart illustrates the distribution of all data collected in two rounds across different time periods, covering most hours of the day.}
      \label{fig-ESM_every_hour}
  \end{minipage}
\end{figure}

We collected a total of 9,002 valid data entries, with 2,356 from the first round and 6,646 from the second.
Table~\ref{tab:Data_description} summarizes the collected data and description. Figure~\ref{fig-ESM_every_hour} shows its distribution across the day, covering nearly all hours and demonstrating the effectiveness of our method in fitting participants' daily routines.

\section{Attention and Current Context}
\label{sec:Current_Context}
This section focuses on exploring the relationship between users' attention and their current context, specifically physical activities, time-related factors, and app usage.

\subsection{Physical Activities}
We record a total of nine types of user physical activities in field study, as summarize in Table~\ref{tab:Behavior}. 
Based on the statistical results, we find that the most common activities are \textit{sitting}, \textit{lying}, \textit{standing still}, and \textit{walking}. 
Among them, \textit{sitting} accounts for over 60\% of the data, with attention states being fairly balanced. This aligns with typical daily life patterns, where people tend to focus on work while \textit{sitting}, but may also take time to relax and rest.
The data from the \textit{lying} state accounts for nearly 15\% of the total, with almost 45\% of users in attention level 1. This trend may be due to the fact that people are generally in a relaxed state when lying down, resulting in a more dispersed attention. Similarly, when users are \textit{walking}, over 45\% fall into level 1. 

We also observed that users tend to have relatively low average attention states while \textit{taking elevator}, \textit{taking transportation}, or \textit{up and down stairs}.
These patterns suggest that although the activities mentioned above involve some physical movement, they generally do not require full concentration to avoid errors. Moreover, people typically do not engage in complex tasks during these moments, which likely contributes to the lower attention states.
In contrast, when users engage in activities such as \textit{cycling/driving} or \textit{running}, their average attention states are significantly higher. This is because \textit{cycling/driving} and \textit{running} are riskier activities that require heightened focus to ensure safety.

We perform a Chi-squared test to investigate these physical activities' relationship with attention states. 
The results reveal a significant difference in the distribution of activities across various attention states ($\chi^2(20, N = 9002) = 274.57, p < .001$). 
Thus, we claim that different physical activities can influence users' attention states.

\begin{table*}
\caption{The table shows statistical values in different physical activities. ``Total'' indicates the data volume for each activity, ``Prop.'' is the proportion, ``Mean'' is the average attention level, ``S.D'' is the standard deviation, and ``Med.'' is the median.}
\label{tab:Behavior}
\small
\resizebox{\textwidth}{!}{
\begin{tabular}{cccccccccc}
    \toprule
    \multirow{2}{*}{\textbf{Activity}} & \multirow{2}{*}{\textbf{Total}} & \multirow{2}{*}{\textbf{Prop.}} & \multicolumn{7}{c}{\textbf{Proportion \& Stats by Attention}}\\
    \cmidrule(lr){4-10}
		& & & \textbf{1} & \textbf{2} & \textbf{3} & \textbf{4} & \textbf{5}  & \textbf{Mean $\pm$ S.D} & \textbf{Med.} \\
    \midrule
    Sitting               & 5606 & \textbf{62.23\%} & 24.88\% & 18.40\%  & 22.15\% & 18.61\% & 15.95\% & 2.67 $\pm$ 1.47 & 3.00 \\
    Lying                 & 1234 & 13.70\%  & \textbf{44.98\%} & 14.59\% & 14.99\% & 12.32\% & 13.13\% & 2.34 $\pm$ 1.47 & 2.00 \\
    Standing sill         & 862  & 9.57\%  & 38.05\% & 19.49\% & 16.24\% & 11.48\% & 14.73\% & 2.45 $\pm$ 1.46 & 2.00 \\
    Walking               & 776  & 8.61\%  & \textbf{45.75\%} & 13.79\% & 16.62\% & 13.79\% & 10.05\% & 2.29 $\pm$ 1.41 & 2.00 \\
    Taking elevator       & 136  & 1.51\%  & 58.09\% & 14.71\% & 13.24\% & 6.62\%  & 7.35\%  & \underline{1.90} $\pm$ 1.28  & \underline{1.00} \\
    Cycling/driving       & 128  & 1.42\%  & 33.59\% & 6.25\%  & 10.94\% & 10.94\% & 38.28\% & \textbf{3.14} $\pm$ 1.75 & 3.00 \\
    Taking transportation & 120  & 1.33\%  & 55.83\% &  8.33\% & 12.50\%  & 13.33\% & 10.00\%    & \underline{2.13} $\pm$ 1.45 & \underline{1.00}\\
    Up and down stairs    & 101  & 1.12\%  & 62.38\% & 11.88\% & 11.88\% & 6.93\%  & 6.93\%  & \underline{1.84} $\pm$ 1.28 & \underline{1.00}\\
    Running               & 45   & 0.50\%   & 33.33\% & 6.67\%  & 6.67\%  & 24.44\% & 28.89\% & \textbf{3.09} $\pm$ 1.69 & 4.00 \\
    \bottomrule
  \end{tabular}}
\end{table*}

\subsection{Time-Related Factors}
We collected and analyzed various time-related factors, including time of day, and weekday status, and now we present a analysis of the relationship between these factors and attention states.

\textbf{Time Periods:} We divided the day into four main periods: morning (6:00-12:00), afternoon (12:00-18:00), evening (18:00-24:00), and late night (0:00-6:00). 
We recorded the time at which participants clicked the ESM surveys and analyzed the corresponding time periods. 
To explore the potential link between attention states and time periods, we conduct a Chi-squared test. 
The results are significant ($\chi^2(12, N = 9002) = 24.42, p < 0.05$), demonstrating a significant association between users' attention states and the time periods they were in.

\textbf{Working Days:} We analyze whether users' responses to ESM surveys occurred on working days. 
A Chi-squared test indicated a significant association between users' attention states and the day of response ($\chi^2(4, N = 9002) = 9.53, p < 0.05$). 
This suggests that users' attention states differ between working days and weekends.

\subsection{App Usage}
We collect 228 unique foreground apps in the field study.
Using Apple's App Store classification method as a reference, we categorize these apps into 22 different categories. 
Table~\ref{tab:App_type} presents the top 10 app categories by proportion, along with the distribution of users' attention states while using these foreground apps.

The first row of the table, ``Home Screen'', indicates that this data was collected when users were not using any apps, accounting for about 41\% of the entire dataset. This shows that our data collection method captures both app engagement and times when users are not using any apps.

We find that users engage most frequently with \textit{Communication} apps (e.g., WeChat), while the largest number of apps falls under the \textit{Utilities} category, with users in our dataset having used a total of 55 different \textit{Utility} apps.

When users engage with \textit{Social} apps, their average attention levels are the lowest. 
This is because social media are typically designed for relaxation and entertainment. 
In contrast, users demonstrate higher attention levels (above 3) when using \textit{Productivity} apps (e.g., Pomodoro ToDo, Samsung Notes) and \textit{Photo \& Video} apps (e.g., Camera, Smile GIF). 
This is because \textit{Productivity} apps, which are linked to work tasks and goal management, encourage users to focus on completing tasks, while \textit{Photo \& Video} apps, often involving creative activities such as taking photos and editing videos, immerse users in content creation, helping them maintain focus.

We also notice that the standard deviation of attention states is the highest when users are using \textit{Music} apps. 
This is because these apps are typically used in a variety of scenarios, including leisure, work, exercise, and more. 
The user's attention states may be influenced by other contexts. 
In addition, different types of music (e.g., classical, rock) with varying styles and rhythms, can have different impacts on users.

These results show that attention states vary across different foreground apps.  By analyzing the app types users engage with, we can infer the tasks or activities they are performing, helping us gain insights into their attention states at a given moment.

\begin{table*}
\caption{The table presents the foreground app types, along with other statistical values. ``Num.'' indicates the number of apps in each category.}
\label{tab:App_type}
\centering
\small
\begin{tabular}{ccccccccccc}
\toprule
\multirow{2}{*}{\textbf{Category}} & \multirow{2}{*}{\textbf{Total}} & \multirow{2}{*}{\textbf{Num.}} & \multirow{2}{*}{\textbf{Prop.}}
& \multicolumn{7}{c}{\textbf{Prop.(\%) \& Stats by Attention}}\\
    \cmidrule(lr){5-11}
		& & & & \textbf{1} & \textbf{2} & \textbf{3} & \textbf{4} & \textbf{5} & \textbf{Mean $\pm$ S.D} & \textbf{Med.} \\
\midrule
Home Screen     & 3702 & - & - & 41.03 & 12.91 & 17.37 & 14.34 & 14.34 & 2.48 $\pm$ 1.49 & 2.00 \\
Communication   & 2235 & 8 & \underline{42.12\%} & 35.79 & 16.96 & 16.42 & 14.27 & 16.55 & 2.59 $\pm$ 1.50 & 2.00 \\
\underline{\textbf{Social}}          & 878 & 10 & 16.55\% & 35.54 & 19.82 & 19.82 & 14.24 & 10.59 & \underline{\textbf{2.45}} $\pm$ 1.37 & 2.00 \\
Entertainment   & 711 & 21 & 13.40\% & 34.04 & 15.89 & 19.13 & 16.32 & 14.63 & 2.62 $\pm$ 1.46 & 3.00 \\
Utilities       & 316 & \underline{55} & 5.96\% & 34.18 & 13.61 & 17.72 & 19.62 & 14.87 & 2.67 $\pm$ 1.48 & 3.00 \\
Shopping        & 274 & 11 & 5.16\% & 35.77 & 18.25 & 17.52 & 14.96 & 13.50 & 2.52 $\pm$ 1.44 & 2.00 \\
Lifestyle       & 177 & 7 & 3.34\% & 26.55 & 15.82 & 19.77 & 13.56 & 24.29 & 2.93 $\pm$ 1.53 & 3.00 \\
Music           & 115 & 5 & 2.17\% & 41.74 & 2.61 & 10.43 & 9.57 & 35.65 & 2.95 $\pm$ \underline{\textbf{1.80}} & 3.00 \\
Education      & 90 & 3 & 1.70\% & 25.56 & 14.44 & 25.56 & 32.22 & 2.22 & 2.71 $\pm$ 1.23 & 3.00 \\
\underline{\textbf{Productivity}}    & 84 & 25 & 1.58\% & 23.81 & 13.10 & 20.24 & 16.67 & 26.19 & \underline{\textbf{3.08}} $\pm$ 1.52 & 3.00 \\
\underline{\textbf{Photo \& Vedio}}  & 75 & 5 & 1.41\% & 14.67 & 22.67 & 24.00 & 22.67 & 16.00 & \underline{\textbf{3.03}} $\pm$ 1.30 & 3.00 \\
\hline
\end{tabular}
\end{table*}

\section{Attention and External Distractions}
\label{sec:handing_behavior}
In our field study, we use notifications as naturally occurring external distractions and collect users’ response behaviors and motivations.
In this section, we combine qualitative coding and quantitative analysis to conduct thematic analysis and descriptive statistics on the open-ended responses about motivations.
Leveraging these valuable data, we explore the relationships between response behaviors and attention, notification types and response motivations, and other related factors, aiming to deepen our understanding of attention and external distractions.

\subsection{Coding Process}
Inspired by prior studies using large language models (LLMs) for qualitative analysis of interview transcripts and forum posts~\cite{bijker2024chatgpt, hamilton2023exploring, naeem2025thematic, li2024safety, li2024double, li2025security}, we explore their feasibility for theme extraction and automated coding in short open-ended responses for the first time. 
Our goal is to improve coding efficiency through human-AI collaboration while maintaining analytical rigor.

First, we randomly select 80\% of the textual data for open and axial coding by two researchers following an inductive thematic analysis process. 
Through discussion, a preliminary category framework is refined.

Meanwhile, we apply the DeepSeek model for latent theme analysis on the same data. 
About 86\% of the model's key themes align with the manual summary, and 56\% of the manual themes appear in the model's output. This trend is consistent with findings from related studies~\cite{hamilton2023exploring}, suggesting that LLMs can play a valuable complementary role in uncovering latent themes.
Based on these results, the final category framework is refined by the researchers, including 46 codes across 16 factors and four high-level categories, as shown in Table~\ref{tab:coding_result}.

Among the four high-level categories, one is ``Notification Content'', which covers all user responses related to the content and type of notifications, such as importance, interest, or further handling needs triggered by specific types of notifications (e.g., alarms, verification codes). 
The other three relate to personal factors: (1) ``Behavioral Patterns'', which include all observable activities reported by users like \textit{entertainment} activities and \textit{cognitive engagement} tasks; (2) ``Individual States'', which include users’ self-reported \textit{levels of busyness} and \textit{emotional/mental states}; and (3) ``Personal-others'', which capture factors such as \textit{personal habits/feelings} that do not fit neatly into the above categories.

After finalizing the category framework, we conduct a reliability check on the aligned coding results from the two researchers. Cohen's Kappa reaches 0.93, indicating high consistency and meeting reliability standards.

Subsequently, we use the DeepSeek model to automatically code the remaining 20\% of data based on the established category framework.
The researchers review and make necessary revisions to the model’s output to ensure coding quality. 
The results show that the model achieves an accuracy of approximately 69.5\%. 
While there remains room for improvement in automatic coding accuracy for short-text scenarios, the approach effectively reduces manual workload and enhances analytical efficiency.

We provide additional details in the Supplemental File.

\begin{table*}
\caption{The table shows our category framework. Numbers after each code indicate the count of notifications assigned to it; note that a single notification may have multiple codes.}
\label{tab:coding_result}
\small
\renewcommand{\arraystretch}{1.02}
\resizebox{\textwidth}{!}{
\begin{tabular}{|l|l|l|}
\hline
\textbf{Categories} & \textbf{Factors} & \textbf{Codes} \\ \hline
 & Not important/not interested 1667 & Direct expression 1254; irrelevant to me 27; ads/marketing 368; etc\\ \cline{2-3} 
 & Important/interested 1602 & \begin{tabular}[c]{@{}l@{}}Direct expression 392; needs reply 892; people mentioned 187;\\ relevant to me 41; etc\end{tabular} \\ \cline{2-3} 
 \multirow{-4}{*}{\begin{tabular}[c]{@{}l@{}}Notification\\ content\\ 3454\end{tabular}} & Requires action 150 & Answer a call 25; dismiss alarm 43;  check verification code 29; etc \\ \cline{2-3} 
 & Check elsewhere/known content 35 &  \\ \hline
 & Entertainment 897 & \begin{tabular}[c]{@{}l@{}}Watching media 346; leisure phone use 140; browsing apps 240; \\playing games 110; etc\end{tabular} \\ \cline{2-3} 
 & Daily life 836 & \begin{tabular}[c]{@{}l@{}}Eating 159; driving/cycling 65; on the way to somewhere 61;\\
 walking 73; washing up 35; viewing sth 77; shopping 41;\\
  exercising 33; waiting 21; etc\end{tabular} \\ \cline{2-3} 
 \multirow{-5}{*}{\begin{tabular}[c]{@{}l@{}}Behavioral\\ patterns\\ 2730\end{tabular}} & Cognitive engagement 786 & \begin{tabular}[c]{@{}l@{}}Working 164; reading 134; doing experiment 87; in a meeting 45; \\in class 61; coding 65; studing 57; reviewing 18; writing 24; etc\end{tabular} \\ \cline{2-3} 
 & Task-switching state 299 & Just done sth 228; just about to do sth 71 \\ \cline{2-3} 
 & Socializing 138 & On a call 23; replying to messages 39; chatting 60; etc\\ \cline{2-3} 
 & Sleep or rest 73 & Sleeping 39; resting 34 \\ \hline
 & Level of busyness 520 & Busy 332; free188 \\ \cline{2-3} 
\multirow{-2}{*}{\begin{tabular}[c]{@{}l@{}}Individual\\state 520\end{tabular}} & Mental/emotional 61 & Mental state 44; emotional state 17 \\ \hline
 & Personal negligence 85 &  \\ \cline{2-3} 
 & Personal habits 16 &  \\ \cline{2-3} 
 & Personal-others 292 & Right timing 33; poor timing 28; notification overload 130\\ 
\hline
\end{tabular}}
\end{table*}

\subsection{Personal Factors and Attention States}
In this section, we take personal-related factors as the entry point to focus on exploring the relationship between different behavioral patterns and attention states at a finer granularity. 
We also aim to understand how these behavioral patterns influence users’ willingness to handle notifications and their response behaviors.

\textbf{Our key findings include:}
\begin{itemize}
    \item Approximately 55\% of users’ motivations for responding notifications are driven by personal factors.
    \item Task complexity affects attention states, which in turn influence notification response behaviors.
    \item During \textit{task switching states}, attention levels are generally low; however, the period \textit{just done sth} serves as a more suitable window for external distractions than the period  \textit{just about to do sth}, with users showing a stronger willingness to interact.
    \item Task proactivity influences interaction willingness. Compared to proactive, cognitively demanding activities, passive cognitive tasks under external constraints (e.g., \textit{in class}) make users more likely to view interruptions as brief opportunities for contextual shifts.
    \item The relaxing and informal nature of activities leads to underestimating actual cognitive engagement.
    \item \textit{Emotional} and \textit{mental states}, as well as the extent of notification backlog, also impact response behaviors.
\end{itemize}

Below, we present the analysis process in detail.

We identify and code a total of 7,198 data points with explicit motivations, and calculate the frequency of each category. 
The results show that 47.99\% of the motivations for responding notifications relate to notification content, while 54.21\% relate to personal factors. 
Further, 45.79\% of notification response motivations are solely related to content, 52.01\% are solely related to personal factors, and 2.2\% are associated with both notification content and personal factors. 
These findings suggest that users’ responses to notifications, as external distractions, are more strongly influenced by personal factors.

Within the behavioral patterns category, we identify a total of 29 fine-grained user behaviors, including four entertainment activities, nine daily life activities, nine cognitively demanding activities, three social activities, two sleep and rest activities, and two state-switching activities.

As shown in Table~\ref{tab:Personal_code}, we calculate the frequency of all coded individual-related factors and the average attention level under different individual states, the overall standard deviation of attention (reflecting overall variability), and the individual-level standard deviation (reflecting within-person variability). 
In addition, we categorize users’ notification response behaviors into three main types—positive response, negative response, and no response—and calculate the proportion of each behavior type under different individual states.

\textbf{Busy or free}: Among all data related to personal factors, 7.2\% are explicitly described by users as indicating a busy or free state. Statistical analysis shows that the average attention level differs by as much as 2.63 between busy and free states, with similarly significant differences observed in notification response behaviors. When users are free, they view 91.49\% of notifications, whereas in a busy state, 82.53\% of notifications remain unattended. These findings suggest that response behavior is closely tied to users’ task engagement: the freer they feel, the more likely they are to check notifications.

\textbf{Task Switching State}: In this state, users exhibit significantly lower attention levels (average 1.85) and a higher tendency to engage with notifications, with only 21.74\% showing no response—over 30\% lower than other activities. 
We further split this state into \textit{just done sth} and \textit{just about to do sth}, finding a 12\% higher positive response rate after task completion.
This suggests users are less sensitive to interruptions when disengaging post-task, whereas they view notifications as more disruptive when preparing for a new task. Thus, the post-task phase may offer a more optimal window for external interruptions.

\textbf{Cognitive Engagement Activities}: This category covers cognitively demanding tasks (e.g., \textit{studying}, \textit{working}) with an average attention level of 3.91, higher than other categories. 
Variability and individual differences are small, and positive response rates are lower. 
Notably, while users report similar attention levels when \textit{in class}, their response rate is higher, possibly because classes, driven by external structure, offer less autonomy. 
Thus, users may view notifications as chances for brief distraction, increasing their willingness to engage.

\textbf{Entertainment Activities}: This category includes activities with clear entertainment purposes, such as watching variety shows and playing games. 
A comparison shows that attention levels during \textbf{playing games} and \textit{leisure phone use} differ by two levels. 
In \textbf{playing games}, the attention level exceeds four, with a 10\% positive response rate. 
This is likely due to the high cognitive engagement required in competitive games. 
In contrast, \textit{leisure phone use} includes data where no explicit entertainment goal is specified, resulting in lower attention levels but a significantly higher positive response rate, over 40\% higher than in similar activities.

\textbf{Daily Life Activities}: This category includes common daily activities, such as \textit{eating} and \textit{shopping}. Users’ attention levels vary significantly due to the differing complexity of these tasks. For instance, attention levels during \textit{exercising} and \textit{driving/cycling} exceed four, while during \textit{waiting}, attention drops below two, with a positive response rate exceeding 57\%.

\textbf{Socializing \& Sleep or Rest Activities}: Social activities involve \textit{chatting} or \textit{on a call}, while sleep or rest refer to \textit{resting}, \textit{sleeping}. 
Although users tend to underestimate their attention during relaxed, informal activities, their notification responses significantly decline as attention shifts to emotional engagement and recovery.

In addition, we observe some interesting patterns. 
Users' emotions and mental states also influence their notification interactions. 
When feeling low or mentally fatigued (e.g., \textit{sleepy}), they are less likely to notice or engage with notifications. 
Additionally, feelings of being overwhelmed by numerous pending notifications lead users to clear them in bulk rather than checking individually.

\begin{table*}
\caption{The table shows attention levels and notification response behaviors link to different personal factors and codes. ``Pos.'' indicates positive responses (``click to view''), ``Neg.'' indicates negative responses (``swipe to clear/clear notification bar'', ``swipe up to cancel when popping up''), and ``No'' indicates no response (``ignore'', ``didn’t notice''). ``Adjust notification settings'' is excluded due to low occurrence. ``Avg'' is average attention, ``All'' is overall standard deviation, and ``Ind'' is average individual standard deviation.}
\label{tab:Personal_code}
\small
\renewcommand{\arraystretch}{1.05}
\resizebox{\textwidth}{!}{
\begin{tabular}{||cccccl||cccccl||}
\toprule
\textbf{Factor/Code} & \textbf{Pos} & \textbf{Neg} & \textbf{No} & \textbf{Avg} & \textbf{{All, Ind}} & \textbf{Factor/Code} & \textbf{Pos} & \textbf{Neg} & \textbf{No} & \textbf{Avg} & \textbf{{All, Ind}} \\ \hline\hline
Busy & 2.71 & 14.46 & 82.53 & \underline{4.06} & 1.16, 1.07 & \textit{\textbf{Cog. Eng.}} & 11.58 & 22.01 & 65.90 & \underline{3.91} & 1.11, 0.84 \\
Free & 91.49 & 5.85 & 2.66 & \underline{1.43} & 0.75, 0.63 & Working & 4.27 & 20.12 & 75.61 & 3.70 & 0.81, 0.93 \\ \cline{1-6} \cline{1-6}  \cline{1-6}
\textit{\textbf{Task-switching}} & 58.53 & 19.73 & 21.74 & 1.85 & 1.18, 1.08 & Reading & 14.18 & 19.40 & 65.67 & 4.10 & 1.11, 1.15 \\
Just done sth & 61.40 & 21.49 & 17.11 & 1.86 & 1.20, 1.10 & Doing Exp. & 12.64 & 22.99 & 64.37 & 4.11 & 0.99, 0.64 \\
Just about to do & 49.30 & 22.54 & 28.17 & 1.83 & 1.12, 1.00 & Coding & 6.15 & 3.08 & 90.77 & 4.15 & 1.41, 0.56 \\ \cline{1-6}
\textit{\textbf{Entertainment}} & 26.2 & 18.95 & 54.74 & 3.23 & 1.44, 1.11 & \underline{In class} & 26.23 & 24.59 & 47.54 & 3.77 & 0.92, 0.82 \\
Watching media & 24.28 & 21.97 & 53.76 & 3.43 & 1.36, 0.95 & Studing & 3.51 & 33.33 & 63.16 & 3.75 & 1.43, 1.18 \\
Browsing apps & 13.75 & 20.42 & 65.83 & 3.37 & 1.39, 0.89 & In a meeting & 15.56 & 15.56 & 66.67 & 3.93 & 1.16, 1.24 \\
\underline{Leisure phone use} & 65.71 & 10.00 & 24.29 & \underline{2.07} & 1.15, 0.88 & Writing & 20.83 & 25.00 & 54.17 & 3.86 & 1.11, 0.79 \\
\underline{Playing games} & 10.00 & 21.82 & 67.27 & \underline{4.08} & 1.21, 1.04 & Reviewing & 0.00 & 72.22 & 27.78 & 3.67 & 0.84, 1.00 \\ \hline
\textit{\textbf{Daily life}} & 20.33 & 16.63 & 62.20 & 3.14 & 1.39, 1.04 & \textit{\textbf{Socializing}} & 2.90 & 26.81 & 69.57 & 3.31 & 1.26, 1.05 \\
Eating & 23.90 & 15.72 & 59.12 & 2.86 & 1.14, 1.01 & Chatting & 5.00 & 25.00 & 68.33 & 3.05 & 1.28, 1.14 \\
Viewing sth & 7.79 & 9.09 & 83.12 & 3.26 & 1.29, 1.30 & On a call & 0.00 & 26.09 & 73.91 & 2.89 & 1.41, 1.60 \\
Walking & 35.62 & 8.22 & 56.16 & 2.59 & 1.45, 1.30 & Reply Msgs. & 0.00 & 25.64 & 74.36 & 3.87 & 1.10, 1.16 \\ \cline{7-12} 
\underline{Driving/cycling} & 4.62 & 16.92 & 78.46 & \underline{4.26} & 1.28, 0.86 & Mental state & 4.55 & 22.73 & 70.45 & 2.05 & 1.33, 1.08 \\
On the way to & 22.95 & 18.03 & 57.38 & 2.82 & 1.28, 1.13 & Emo. state & 0.00 & 29.41 & 64.71 & 2.67 & 1.24, 1.29 \\ \cline{7-12} 
Shopping & 9.76 & 31.71 & 58.54 & 3.66 & 1.37, 0.83 & Pers. Neglig. & 0.00 & 5.88 & 94.12 & 2.90 & 1.50, 1.31 \\
Washing & 5.71 & 8.57 & 80.00 & 3.00 & 1.24, 1.22 & Pers. habits & 6.25 & 56.25 & 37.50 & 3.38 & 1.45, 1.63 \\
\underline{Exercising} & 12.12 & 12.12 & 75.76 & \underline{4.24} & 1.20, 1.16 & Poor timing & 0.00 & 57.14 & 7.14 & 3.25 & 1.00, 1.10 \\
\underline{Waiting} & 57.14 & 19.05 & 23.81 & \underline{1.81} & 1.12, 0.74 & Right timing & 87.88 & 12.12 & 0.00 & 1.33 & 0.85, 0.98 \\ \cline{1-6} \cline{1-6}  \cline{1-6}
\textit{\textbf{Sleep or rest}} & 15.07 & 21.92 & 61.64 & 2.77 & 1.40, 1.16 & Notif. OL & 0.00 & 89.23 & 5.38 & 1.52 & 0.99, 1.34 \\ 
\toprule
\end{tabular}}
\end{table*}

\subsection{Response Behaviors and Attention States}
In the previous section, we analyzed users' fine-grained behavioral activities in notification response motivations, and identified a close link between task characteristics and attention states. 
This suggests that users' responses to notifications are shaped by their current tasks, or more precisely, by their attention state. 
In this section, we further examine the relationship between response behaviors and attention states.

Unlike prior work that focused only on whether notifications were viewed or dismissed, we captured six distinct actions, covering active/passive ignoring, system-level settings changes, and varied dismissal behaviors. 
This fine-grained categorization helps reveal the differing user motivations behind similar outward outcomes.

We run a chi-square test and find significant differences in response behaviors across attention levels ($\chi^2(20, N = 9002) = 1278.92, p < 0.001$), confirming that attention shapes notification response behaviors and supporting the use of such interactions as indicators of attention states.

We further calculated the average attention levels in the valid data under different response behaviors and analyzed the underlying motivation types in the coded data. 
The proportions of notification-related and personal-related factors are shown in Table~\ref{tab:Handling_Behavior}.

For the same negative response behavior, the average attention level for ``swipe up to cancel when popping up'' is 0.6 levels higher than ``swipe to clear/clear notification bar''.
12.5\% of motivations for `swipe up to cancel when popping up'' are related to \textit{watching media}, and 7.5\% to \textit{using app}, suggesting that this behavior occurs when users are engaged with the phone interface and find pop-up notifications distracting.
On the other hand, 30\% of the motivations for ``swipe to clear/clear notification bar'' are due to \textit{notification overload}, and 6.5\% are for \textit{just done sth}. 
This aligns with the previous section, where users in a task-switching state show lower attention and are more likely to handle notifications. However, excessive accumulation reduces their desire to check them, leading to direct clearing.

Although both ``ignore'' and ``didn’t notice'' are both non-responses from a system perspective, they differ in user intent: ``ignore'' reflects active disregard, while ``didn’t notice'' signals passive unawareness.
12\% of ``ignore'' cases are due to being \textit{busy}, compared to 14.8\% for ``didn’t notice''. The average attention for ``didn’t notice'' is nearly one level higher, suggesting users actively ignore when moderately focused but as attention increases, users shift from actively blocking interruptions to passively tuning them out.
In terms of behavioral patterns, 14.2\% of ``ignore'' cases occur during \textit{entertainment} activities, while 24.2\% of ``didn't notice'' cases happen during \textit{cognitive engagement}. 
This further indicates that active disregard is more likely in situations of moderate-to-high attention, where users are still able to perceive external distractions, while passive unawareness is associated with deep \textit{cognitive engagement}, leading to a significantly higher threshold for noticing distractions.

We find that attention is lowest when users ``clicking to view''—averaging 1.75 levels below ``ignore''—suggesting positive responses tend to occur when attention is more scattered. Notably, ``click to view'' is the only behavior mainly driven by notification factors, while others are mainly influenced by personal motivations. 
Besides, as attention increases, behavior becomes more driven by personal factors; when users report ``didn't notice'', personal factors account for as much as 99.48\%.
This suggests high-focus states anchor responses to ongoing tasks, while low-focus states heighten sensitivity to the nature of external distractions.

``Adjust notification settings'' is a strategic behavior not tracked by NotificationListenerService class, but we include it to better understand users' patterns. 
Among its motivations, 29.4\% cite ``poor timing'' and 20.5\% cite ``notification overload'', suggesting that users make system-level changes when experiencing ongoing disruption.

\begin{table*}
\caption{The table shows response behaviors with attention levels and motivations. ``Mean $\pm$ S.D'' gives the average attention and standard deviation. ``Notif. Factors'' and ``Pers. Factors'' show the share of notification- and personal-related motivations.}
\label{tab:Handling_Behavior}
\small
\begin{tabular}{cccccc}
    \toprule
    \textbf{Response Behavior} & \textbf{Total} & \textbf{Prop.} & \textbf{Mean $\pm$ S.D} & \textbf{Notif. Factors} & \textbf{Pers. Factors}\\
    \hline
    Click to view
    & 3617 & 40.18\% & \underline{2.08} $\pm$ 1.30 & 65.54\% & 34.46\%\\
    \hline
    Swipe to clear/clear notification bar
    & 912 & 10.13\% & \underline{2.17} $\pm$ 1.38 & 47.76\% & 52.24\%\\
    Swipe up to cancel when popping up
    & 795 & 8.83\% & 2.77 $\pm$ 1.37 & 35.29\% & 64.71\%\\
    \hline
    Ignore 
    & 2975 & 33.05\% & \textbf{2.93} $\pm$ 1.50 & 41.66\% & 58.34\%\\
    Didn't notice
    & 612 & 6.80\% & \textbf{3.83} $\pm$ 1.38 & 0.52\% & 99.48\%\\
    \hline
    Adjust notification settings
    & 91 & 1.01\% & \underline{2.15} $\pm$ 1.24 & 43.33\% & 56.47\%\\
    \bottomrule
  \end{tabular}
\end{table*}

\subsection{Motivational Differences Across Notification Categories}
As 47.99\% of response motivations are content-related, this prompts us to explore how primary motivations differ across notification categories.

Our key findings are:
\begin{itemize}
    \item Response motivations vary by notification category, with low-novelty notifications mainly influenced by personal factors, while notifications with communicative attributes or high content variability are primarily driven by content relevance.
\end{itemize}

\begin{table*}
\caption{The table shows the response behaviors and influencing factor proportions for major notification types. Here, ``Notif.'' is the proportion of notification-related factors, ``Pers.'' is personal factors. ``Example'' gives typical examples, and ``Main factors \& Prop.'' lists key response motivations and their proportions.}
\label{tab:Notif_Type}
\small
\resizebox{\textwidth}{!}{
\begin{tabular}{lllllllll}
\toprule
\textbf{Type} & \textbf{Prop.} & \textbf{Pos} & \textbf{Neg} & \textbf{No} & \textbf{Notif.} & \textbf{Pers.} & \textbf{Example} & \textbf{Main factors \& Prop.} \\ \hline
Comm. & \underline{42.8} & 52.8 & 12.5 & 34.1 & \underline{52.8} & 47.2 & \begin{tabular}[c]{@{}l@{}}Wechat;\\  QQ\end{tabular} & \begin{tabular}[c]{@{}l@{}}Needs reply 26\%; People mentioned 5.3\%;\\ Not important/interested 9.7\%; \\ Important/interested 5.8\%;\end{tabular} \\ \hline
System & 15.0 & 26.6 & 27.2 & 44.1 & 30.4 & \underline{69.6} & \begin{tabular}[c]{@{}l@{}}System; \\ Theme\end{tabular} & \begin{tabular}[c]{@{}l@{}}Just done sth 7.6\%; Using APP 6.0\%;\\ Watching media 7.3\%;\end{tabular} \\ \hline
Social & 10.3 & 28.1 & 17.2 & 54.6 & \underline{54.5} & 45.5 & \begin{tabular}[c]{@{}l@{}}Weibo; \\ RedNote\end{tabular} & \begin{tabular}[c]{@{}l@{}}Not important/interested 29.9\%; \\ Important/interested 13.1\%; \\ Ads/marketing 5.4\%\end{tabular} \\ \hline
Utilities & 9.1 & 18.6 & 32.4 & 47.7 & 45.3 & \underline{54.7} & \begin{tabular}[c]{@{}l@{}}Alarm; \\ App store\end{tabular} & \begin{tabular}[c]{@{}l@{}}Busy 11.3\%; Watching media 4.6\%; \\ Notif. OL 4.1\%; Just done sth 3.3\%\end{tabular} \\ \hline
Shopping & 3.7 & 6.5 & 24.7 & 68.4 & \underline{65.0} & 35.0 & \begin{tabular}[c]{@{}l@{}}JD; \\ Taobao\end{tabular} & \begin{tabular}[c]{@{}l@{}}Ads/marketing 43\%; \\ Not important/interested 19.3\%\end{tabular} \\ \hline
Health \& Fit. & 3.6 & 33.2 & 16.4 & 50.4 & 9.2 & \underline{90.8} & \begin{tabular}[c]{@{}l@{}}Healthy \\ Phone Use\end{tabular} & \begin{tabular}[c]{@{}l@{}}Just done sth 15.2\%; Using APP 7.2\%; \\ Reading 6.1\%\end{tabular} \\ 
\bottomrule
\end{tabular}}
\end{table*}

Table~\ref{tab:Notif_Type} summarizes response behaviors' distribution and motivations for the six most common notification categories in the study.
Results show that for \textit{Communication}, \textit{Social}, and \textit{Shopping} notifications, user motivations are more content-related, while \textit{System}, \textit{Utilities}, and \textit{Health \& Fitness} are more shaped by personal factors.

We attribute this difference to the typically low novelty and variability of \textit{System}, \textit{Utilities}, and \textit{Health \& Fitness} notifications, which mainly serve functional or reminder roles, such as system updates or step counts, and are less tied to users’ work, social needs, or interests.
Thus, responding to them depends more on user availability than content appeal or urgency.
In contrast, \textit{Shopping} and \textit{Social} notifications are mostly platform-driven to grab attention and encourage interaction or spending, often novel and highly variable but less relevant to user.
As a result, many users perceive them as marketing messages with low importance and interest.
Meanwhile, instant messages, key for social connection, are handled based on message importance, social relevance, and response obligation — with a positive response rate nearing 53\%, far above others, reflecting their higher user priority.

\subsection{Response Time}
In previous studies, response time has been regarded as a key indicator of user behavior~\cite{iqbal2008effects, 0How, chang2019think}. Here, we examine the relationship between users’ response time to external distractions and their attention by calculating the time difference between when they receive and click on an ESM survey.

Table \ref{tab:respond_time} presents the average response time, median, and interquartile range (Q1 and Q3) across different levels of attention. As the attention level increases, response time also increases, particularly from level 1 to level 2, where both the average and median nearly double. However, due to the possibility of ESM surveys remaining in the notification bar for long periods, some surveys are clicked much later, leading to extreme values that inflate the mean. In contrast, the median and quartiles remain concentrated around a specific range, indicating that the majority of response times are not as heavily affected by these outliers.

Given the skewed distribution of response times~\cite{lo2015transform}, we employed a Generalized Linear Mixed Model (GLMM) to analyze the relationship between attention and response time, as GLMM handles skewed data more effectively, avoiding the biases of linear regression. 
We treat attention as a fixed effect and included a quadratic term to capture potential non-linear trends. The results are presented in Table \ref{tab:GLMM}. 

The results show that for each unit increase in attention, response time increases by 48.739 seconds ($ p < 0.01 $), indicating a significant positive effect of attention on response time. 
Moreover, the coefficient for the quadratic term is -1.316 ($ p = 0.032 $), suggesting that the rate of increase in response time slows as attention rises.
The ``Group Var'' represents the degree of variability between different individuals. Our results show that there is notable variability in response times across individuals. We include the random effect coefficients for all individuals in the Appendix~\ref{sec:appendix_GLMM}.

\begin{figure}[h]
  \centering
  \begin{minipage}[c]{0.37\textwidth}
    \centering
    \captionof{table}{The table shows response time statistics. Despite outliers affecting the mean, response time at level 1 is significantly lower than at other levels.}
    \label{tab:respond_time}
    \small
    \resizebox{\textwidth}{!}{
    \begin{tabular}{ccccc}
    \toprule
    \textbf{Attention} & \textbf{Mean} & \textbf{Med.} & \textbf{Q1} & \textbf{Q3}\\
    \midrule
    1 & 138.78 & 24 & 11 & 101\\
    2 & 219.64 & 47 & 13 & 155\\
    3 & 215.73 & 40 & 12 & 178\\
    4 & 279.45 & 43 & 12 & 165\\
    5 & 285.66 & 62 & 14 & 207\\
    \bottomrule
  \end{tabular}}
  \end{minipage}
  \hfill
    \begin{minipage}[c]{0.6\textwidth}
        \centering
\captionof{table}{The table shows the GLMM regression results. \textit{Coef.} represents the fixed variable's effect on response time (positive values indicate an increase). \textit{P} indicates the statistical significance of the coefficients. \textit{95\% CI}: Confidence intervals for all coefficients exclude zero, confirming significance at the 95\% confidence level.}
\label{tab:GLMM}
\small
\resizebox{\textwidth}{!}{
    \begin{tabular}{cccccc}
    \toprule
     & \textbf{Coef.} & \textbf{Std.Err.} & \textbf{z} & \textbf{P} & \textbf{95\% CI}\\
    \midrule
    $Intercept$ & 26.49 & 5.91 & 4.48 & <.001 & (14.91,38.07)\\
    $Attention$ & 18.82 & 3.63 & 5.18 & <.001 & (11.7,25.95)\\
    $Attention^2$ & -1.32 & 0.62 & -2.14 & 0.032 & (-2.52,-0.11)\\
    $Group Var$ & 648.27 & 2.03 & & & \\
    \bottomrule
  \end{tabular}}
    \end{minipage}
\end{figure}

\section{Attention State Prediction}
\label{sec:attention_predict}
Existing studies on users' subjective mental states often rely on intrusive or wearable devices or extensive personalized historical data, limiting their generalization and real-world deployment.
To address this, we find that users’ contextual information and responses to external stimuli are strongly associated with attention states. 
Based on this insight, we propose \textbf{AttenTrack}, a multi-feature fusion model that jointly models context and response features to effectively predict users’ attention levels without requiring personalized history.

\subsection{Model Design}
\begin{figure}[h]
  \centering
  \includegraphics[width=\linewidth]{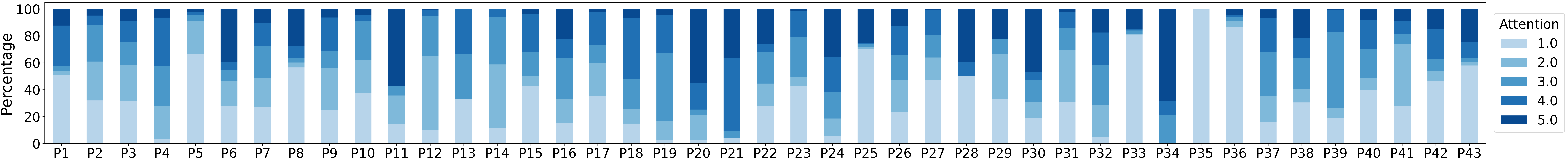}
  \caption{The figure shows the distribution of attention levels among all participants.}
  \label{fig:35UserAttention}
\end{figure}

we analyze the variations in users' self-reported attention scores (see Fig. \ref{fig:35UserAttention}). 
The results reveal substantial individual differences in rating preferences: some users tend to rate their attention levels higher, while others prefer lower values.
This finding pose a challenge for model classification, requiring the definition of unified thresholds that apply across the majority of users.

Through the previous analysis, we observe that:

Users generally exhibit attention levels of $\geq$3 during complex tasks (e.g., ``busy'', ``driving/riding'', ``cognitive engagement'') and levels of $<$ 3 during relaxed or idle states (e.g., ``free'', ``task switching state''). 
Based on this, we set level 3 as the threshold to classify attention states into ``less focused'' and ``more focused'', and build the \textbf{AttenTrack\,I} model accordingly. 
The model aims to intelligently detect declining attention trends and apply to attention maintenance and intervention scenarios, thereby enhancing collaborative efficiency in human-computer interaction.

Further, we find that despite individual rating biases, users show consistency in self-perception of being ``completely unfocused'' (level 1). Once attention levels exceed 1, users generally perceive their state as distinct from being ``completely unfocused''.
This insight leads to the development of the \textbf{AttenTrack\,II}, which distinguishes between ``completely unfocused'' and ``somewhat focused'' states. 
This model is suitable for scenarios where optimizing push timing is crucial, such as advertising and marketing communications. 
When users are in a completely unfocused state, pushing information aimed at stimulating user engagement typically results in higher response rates and participation, thus improving the effectiveness of information delivery.

Additionally, we explore the feasibility of multi-level attention prediction for unknown users, categorizing attention level 1 as ``low focus'', levels 2-3 as ``medium focus'', and levels 4-5 as ``high focus''. 
Based on this, we develop the three-class model \textbf{AttenTrack\,III} for more granular attention classification tasks.

The three \textbf{AttenTrack} models are summarized as follows:

\textbf{AttenTrack\,I}: Distinguishes between ``less focused'' ($<$ 3) and ``more focused'' ($\geq$3).

\textbf{AttenTrack\,II}: Distinguishes between ``completely unfocused'' (=1) and ``somewhat focused'' ($>$ 1).

\textbf{AttenTrack\,III}: Predicts three attention levels: ``low'' (1), ``medium'' (2-3), and ``high'' (4-5).

It is worth noting that in the field study, we collect six types of fine-grained notification response behaviors. To facilitate practical system application, we combine ``ignored'' and ``didn't notice'' into ``no response'', aligning the feature design with the system's ability to recognize coarser behaviors.

\begin{table*} [htbp]
  \caption{The table presents the average results and standard deviations of predicting the attention states of all participants when each participant is alternately treated as an unknown user (i.e., test set).}
  \label{tab:Attention_result}
  \small
  \resizebox{\textwidth}{!}{\begin{tabular}{cccccc}
    \toprule
    \textbf{Model} & \textbf{Acc. $\pm$ SD} & \textbf{F1 $\pm$ SD} & \textbf{Pre. $\pm$ SD} & \textbf{Re. $\pm$ SD}& \textbf{AUC $\pm$ SD}\\
    \midrule
    Baseline I & 49.11\% $\pm$ 0.05& 49.17\% $\pm$ 0.08& 55.40\% $\pm$ 0.18& 46.69\% $\pm$ 0.06 & 49.75\% $\pm$ 0.04\\
    AttenTrack I\_$_{\mathrm{RF}}$ &62.45\% $\pm$ 0.11 &60.73\% $\pm$ 0.15 &69.87\% $\pm$ 0.15 &56.22\% $\pm$ 0.13& 67.59\% $\pm$ 0.13\\
    AttenTrack I\_$_{\mathrm{response}}$\_$_{\mathrm{RF}}$ &62.58\% $\pm$ 0.14 &61.00\% $\pm$ 0.16 &69.69\% $\pm$ 0.16 &57.85\% $\pm$ 0.20& 68.01\% $\pm$ 0.14\\
    AttenTrack I\_$_{\mathrm{factor}}$\_$_{\mathrm{RF}}$ &77.01\% $\pm$ 0.12 &83.21\% $\pm$ 0.11 &81.45\% $\pm$ 0.12 & 86.95\% $\pm$ 0.14 & 74.15\% $\pm$ 0.12\\
    AttenTrack I\_$_{\mathrm{GB}}$ & 64.72\% $\pm$ 0.12& 65.29\% $\pm$ 0.12& 69.12\% $\pm$ 0.12& 64.72\% $\pm$ 0.12 & 69.52\% $\pm$ 0.14\\
    AttenTrack I\_$_{\mathrm{response}}$\_$_{\mathrm{GB}}$ & 61.59\% $\pm$ 0.11& 64.88\% $\pm$ 0.10& 74.43\% $\pm$ 0.12& 61.59\% $\pm$ 0.11 & 67.01\% $\pm$ 0.13\\
    AttenTrack I\_$_{\mathrm{factor}}$\_$_{\mathrm{GB}}$ & 74.86\% $\pm$ 0.14& 75.10\% $\pm$ 0.14& 80.28\% $\pm$ 0.10& 74.86\% $\pm$ 0.13 & 79.53\% $\pm$ 0.13\\
    \midrule
    Baseline II & 56.39\% $\pm$ 0.04 & 66.70\% $\pm$ 0.07 & 73.85\% $\pm$ 0.16& 62.79\% $\pm$ 0.03 & 51.46\% $\pm$ 0.07\\
    AttenTrack II\_$_{\mathrm{RF}}$ & 70.52\% $\pm$ 0.10& 80.09\% $\pm$ 0.08& 76.70\% $\pm$ 0.14& 85.84\% $\pm$ 0.07 & 66.30\% $\pm$ 0.12\\
    AttenTrack II\_$_{\mathrm{GB}}$ & 61.84\% $\pm$ 0.11& 65.19\% $\pm$ 0.10& 74.40\% $\pm$ 0.12& 61.84\% $\pm$ 0.11 & 67.76\% $\pm$ 0.13\\
    \midrule
    Baseline III & 32.26\% $\pm$ 0.05 & 29.99\% $\pm$ 0.05 & 32.50\% $\pm$ 0.04& 33.06\% $\pm$ 0.05 & 49.65\% $\pm$ 0.04 \\
   AttenTrack III\_$_{\mathrm{RF}}$ & 41.48\% $\pm$ 0.12& 38.45\% $\pm$ 0.09& 41.05\% $\pm$ 0.08& 43.05\% $\pm$ 0.08 & 61.08\% $\pm$ 0.08 \\
   AttenTrack III\_$_{\mathrm{GB}}$ & 44.74\% $\pm$ 0.12& 46.22\% $\pm$ 0.13& 52.58\% $\pm$ 0.14& 44.74\% $\pm$ 0.12 & 61.80\% $\pm$ 0.08 \\
    \bottomrule
  \end{tabular}}
\end{table*}

\subsection{Model Evaluation}
Given that attention state detection commonly faces cold start issues in real-world deployment, we employ a Leave-One-User-Out strategy to simulate scenarios where a new user arrives or user data is sparse, thereby evaluating the model's generalization ability to unknown users. 
The \textbf{AttenTrack} is trained using Random Forest (RF) and Gradient Boosting (GB), with random prediction as the Baseline for comparison. The results are shown in Table~\ref{tab:Attention_result}, and detailed experimental procedures are provided in Appendix~\ref{sec:appendix_model}.

In \textbf{AttenTrack\,I} (``less focused'' vs. ``more focused''), our GB model outperforms with an accuracy of 64.72\% and a precision of 69.12\% for unknown users.
The recall improves by approximately 18\% over the Baseline1, and the AUC increases by about 20\%.
Furthermore, in AttenTrack I\_$_{\mathrm{response}}$, we compare the performance of fine-grained notification response behavior features with coarse-grained, system-recognizable features (i.e., AttenTrack I). 
We find that the difference is minimal, and even AttenTrack I\_$_{\mathrm{GB}}$ outperforms AttenTrack I\_$_{\mathrm{response}}$\_$_{\mathrm{GB}}$. 
This indicates that AttenTrack can maintain strong predictive performance with simplified feature design, which is beneficial for deployment in real-world systems.

Additionally, in the AttenTrack I\_$_{\mathrm{factor}}$, we introduce fine-grained personal factors extracted through qualitative coding from the open-ended question as additional feature inputs.
This boosts the RF model's F1 score to 83.21\% and recall to 86.95\%, while the GB model's AUC increases to 79.53\%, showing significant improvement over AttenTrack I. 
This validates that, despite the significant individual differences that make cold-start problems a challenge in this field, incorporating richer, fine-grained contextual information enables the model to overcome individualization limitations and more accurately capture common attention trends across users.

In the \textbf{AttenTrack\,II} (``completely unfocused'' vs. ``somewhat focused''), the RF model performs better, achieving an average F1 score of 80.09\% and a recall rate of 85.84\% for unknown users, an improvement of over 23\% compared to the Baseline2. 
This demonstrates that our model has strong generalizability and sensitivity in identifying users' completely unfocused states.

In the \textbf{AttenTrack\,III} (low-medium-high focus), the GB model achieves an average F1 score of 46.22\% and a precision of 52.58\% for unknown users, with an AUC improvement of over 20\% compared to the Baseline3. 
This demonstrates the feasibility of AttenTrack for multi-level attention state classification in unknown users, laying the foundation for more generalized and fine-grained subjective state perception.

\begin{figure}[h]
  \centering
    \begin{minipage}[c]{0.59\textwidth}
    \centering
      \includegraphics[width=\textwidth]{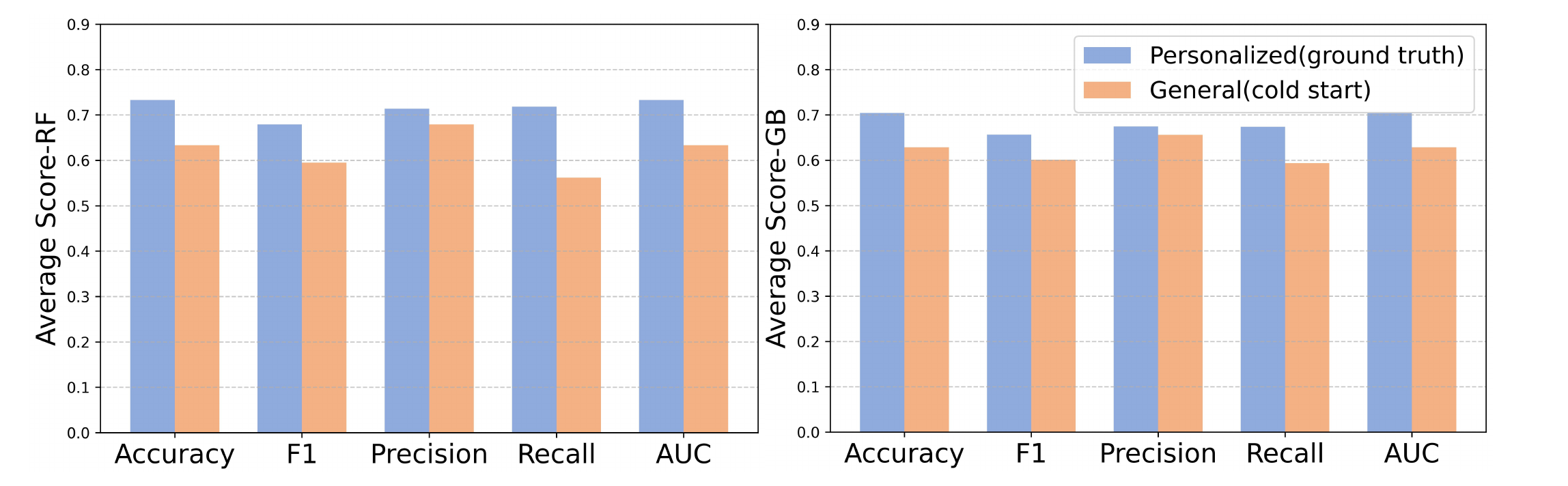}
      \captionof{figure}{The figure presents a comparison between the personalized models (with ground truth) and the general model (cold-start), with the results of RF shown on the left and GB on the right. In this experiment, we sort each user’s data chronologically, using the first 70\% to train the personalized model, while the remaining 30\% serve as the common test set for both the personalized and general models.}
      \label{fig:RFBG_User_30Test}
  \end{minipage}
  \hfill 
  \begin{minipage}[c]{0.23\textwidth}
    \centering
      \includegraphics[width=\textwidth]{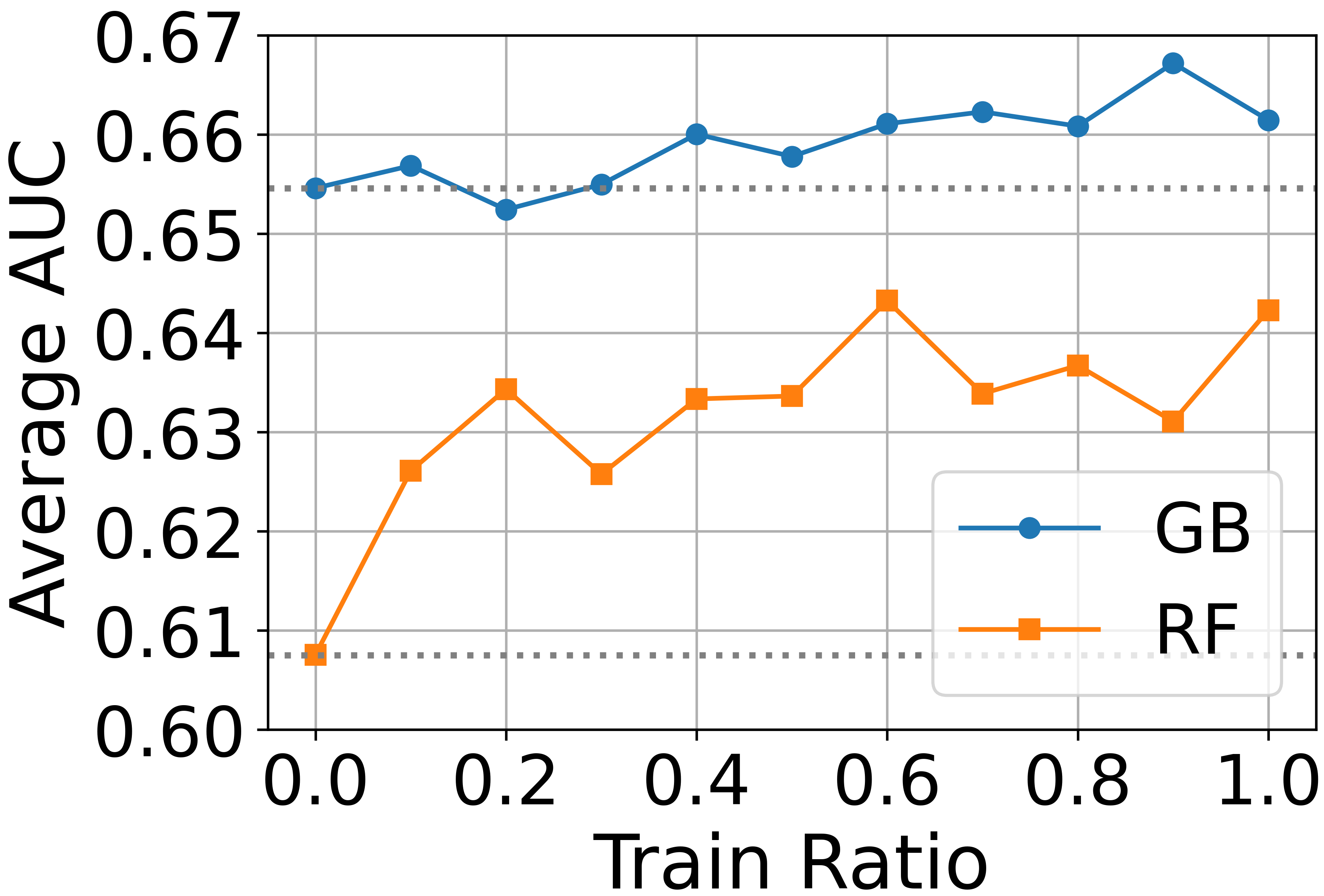}
      \captionof{figure}{The figure shows how adding personal data affects model AUC, using the last 20\% of each user’s data for testing and incrementally adding portions of the first 80\% for training.}
      \label{fig:RFGB}
  \end{minipage}
    \hfill 
  \begin{minipage}[c]{0.14\textwidth}
    \centering
      \includegraphics[width=\textwidth]{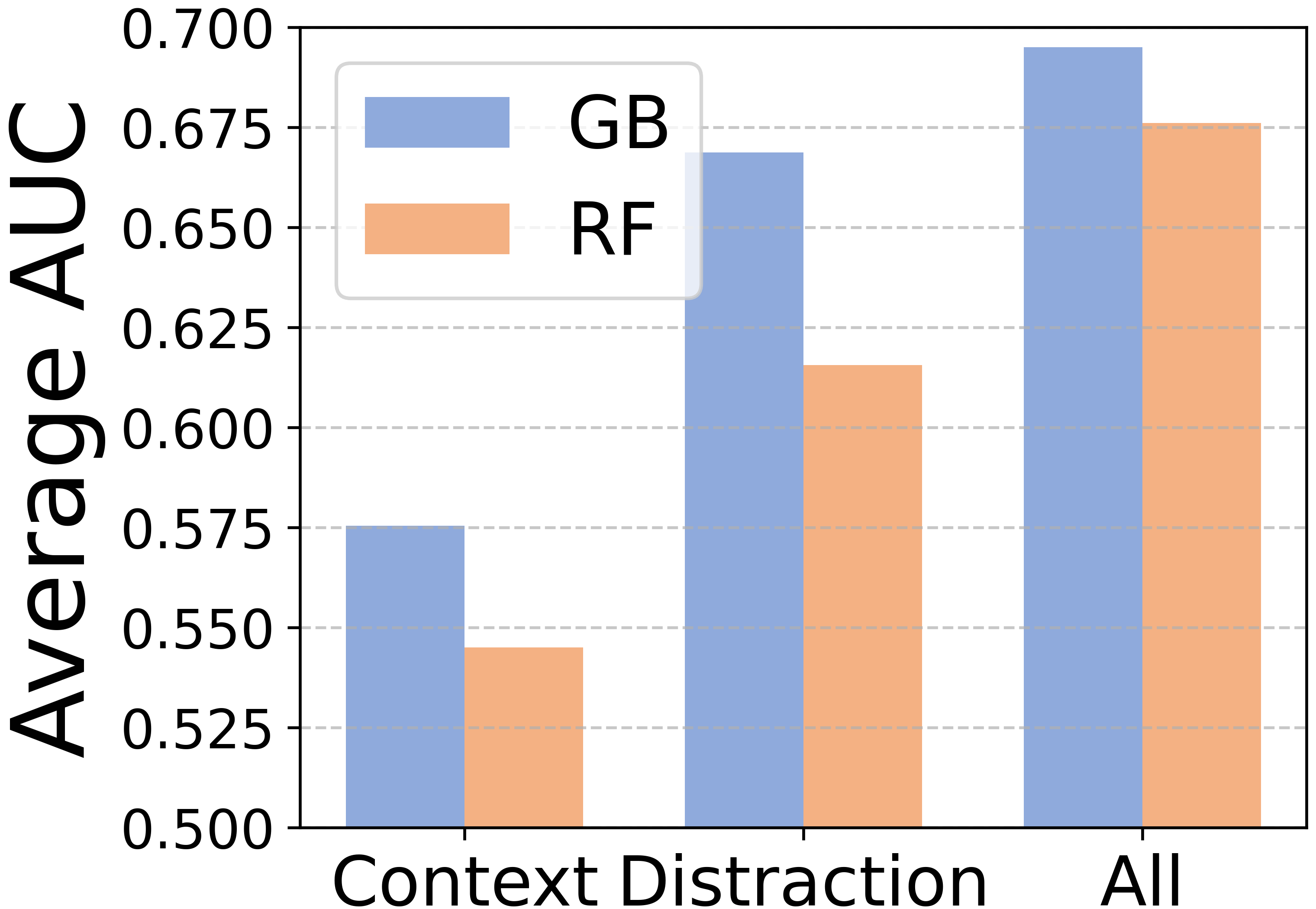}
      \captionof{figure}{The figure shows the average AUC when utilizing only context features, only external distraction features, and all features.}
      \label{fig:feature_combination}
  \end{minipage}
\end{figure}

\begin{figure}[h]
  \centering
  \begin{minipage}[c]{0.33\textwidth}
    \centering
      \includegraphics[width=\textwidth]{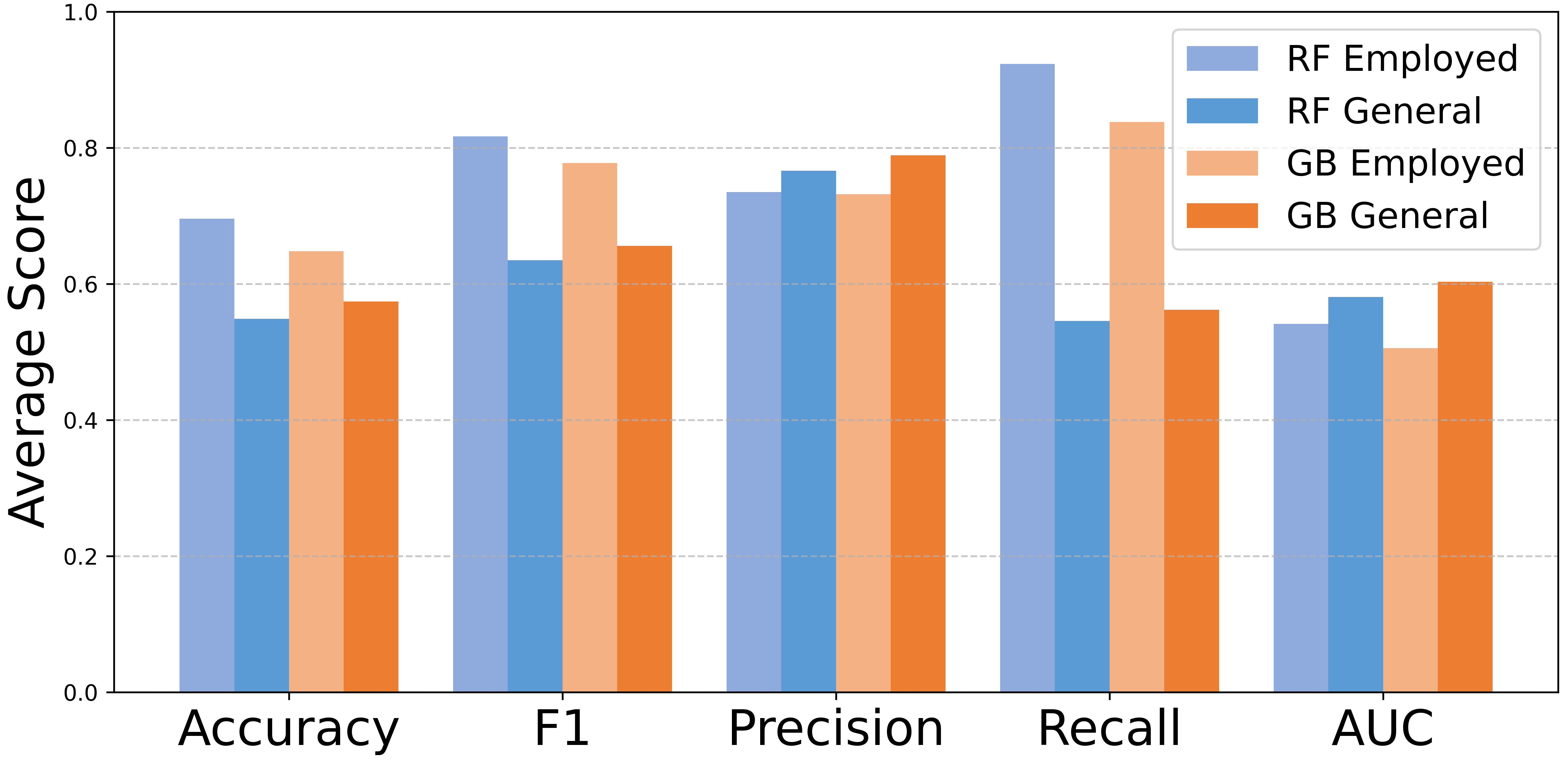}
      \captionof{figure}{The figure shows the average cold-start performance of the group model and the general model for  employed users.}
      \label{fig:Work_Comparison}
  \end{minipage}
  \hfill 
  \begin{minipage}[c]{0.65\textwidth}
    \centering
      \includegraphics[width=\textwidth]{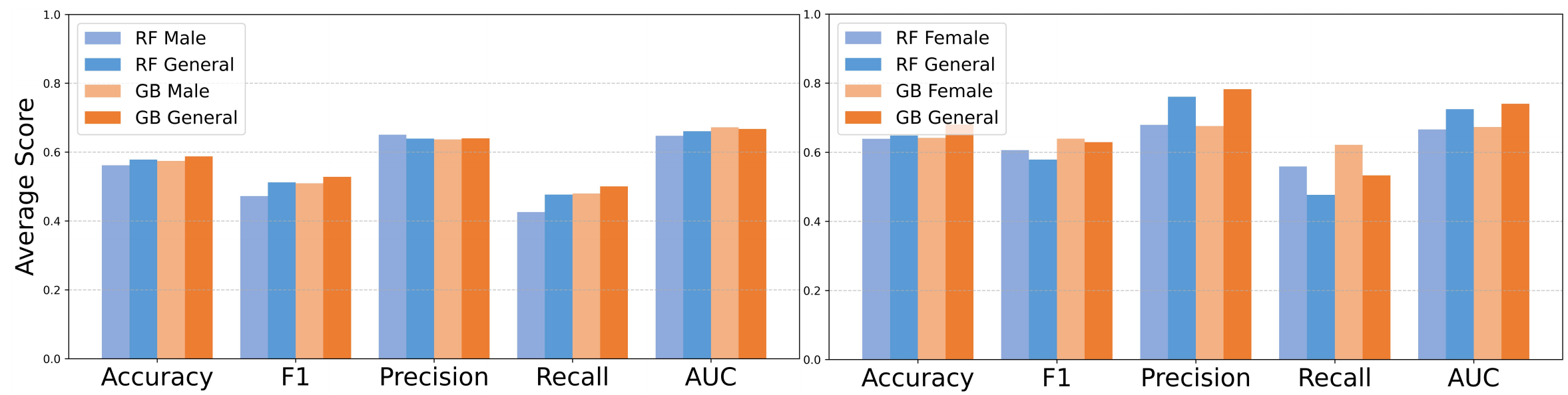}
      \captionof{figure}{The figure shows the average cold-start performance of the group models and the general model across different genders. The left panel presents the group model for male users, while the right panel shows the group model for female users.}
      \label{fig:woman_man}
  \end{minipage}
\end{figure}

\subsection{Personalization Evaluation}
In this section, we conduct further experiments to investigate the impact of different user characteristics on model performance. It is important to note that all models are constructed based on the \textbf{AttenTrack\,I} architecture.

First, we investigate the impact of personal user data on model performance. 
For each user, we fix the last 30\% of their data as the test set and use the first 70\% (averaging around 14 days) for training. 
We compare a personalized model trained on the user’s own data (with ground truth) against a general model trained on data from other users (cold start), as shown in Figure~\ref{fig:RFBG_User_30Test}. 
The results show that, after collecting 14 days of data on average, the personalized model built by AttenTrack achieves nearly a 10\% improvement in accuracy over the general model.
Furthermore, we fix the last 20\% of each user’s data as the test set and gradually incorporate different proportions of their remaining 80\% personal data into the general model’s training set.
As shown in Figure~\ref{fig:RFGB}, model performance improves progressively as more user-specific data are incorporated. 
This suggests that AttenTrack can achieve further performance gains in real-world deployment by incrementally collecting more personal data.

We further compare the impact of different feature combinations on model performance. 
Specifically, we evaluate models using only contextual features, only external distraction features, and all features combined, and report their average AUC (see Figure~\ref{fig:feature_combination}). 
The results show that models using a single feature category perform worse than the combined model, indicating that integrating contextual and external distraction features enables more effective prediction of users’ attention states.

Next, we investigate the impact of user group characteristics on model performance. 
Specifically, we build a group model for employed users and compare it with the general model. 
As shown in Figure~\ref{fig:Work_Comparison}, the group model shows significant performance gains, with improvements in accuracy, F1 score, and recall, the latter increasing by more than 20\%. 
These findings suggest that tailoring models to specific occupational groups can effectively enhance AttenTrack’s cold-start performance.

Furthermore, we explore the impact of gender differences within the student user group. 
We build separate group models for each gender and compare them with the general model. 
As shown in Figure~\ref{fig:woman_man}, the gender-specific models do not yield significant performance improvements, indicating that gender differences do not have a notable impact on attention state prediction.

\section{Discussion}
\label{sec:discussion}
To better adapt to diverse attention management scenarios, we conduct an in-depth analysis of the intrinsic relationship between users’ contextual information, their response patterns to external distractions, and their attention states. 
Based on this, we design three AttenTrack models that enable cold-start prediction of users’ attention states without requiring personalized prior knowledge. 
This opens up new possibilities for personalized assistance and intervention in intelligent interactive systems.
Next, we discuss the lessons learned from our analysis, examine the application potential of AttenTrack in different scenarios, and further summarize its limitations and future work.

\subsection{Lessons Learned}
Mobile notifications, as a common external source of distraction, not only help us infer users' attention states but also provide deeper insights into their response patterns to notifications. 
We find that notifications with novel and varied content are more likely to drive user interaction, and users’ sensitivity to notification content increases as their attention decreases. 
Therefore, from a system design perspective, non-urgent notifications intended to engage users are better suited for times when users are not focused on a task, such as during task-switching phases. 
In contrast, responses to functional notifications are more influenced by users' internal states, making them better suited as external distractions for inferring attention fluctuations.

Although AttenTrack effectively predicts attention states without prior information, further personalization evaluation shows that constructing fine-grained group models leads to better detection performance. 
We find that user groups with similar lifestyles exhibit more comparable attention patterns, while factors like gender have a relatively limited impact. 
Additionally, when the system receives even a small amount of real user feedback, model performance improves significantly. 
This indicates that AttenTrack has strong scalability, and future deployments can further enhance application effectiveness through lightweight personalized calibration.

\subsection{Case Analysis}
To promote the practical application of attention-aware technology, we explore the value of three AttenTrack models in different scenarios.

\textbf{Attention maintenance scenario.}
\textbf{AttenTrack I} focuses on distinguishing users' ``less focuse'' and ``more focused'' states, making it suitable for educational assistance and group collaboration platforms, where it helps users maintain focus and improve efficiency. 
For example, in online learning platforms, the system can dynamically adjust the teaching pace based on the student's current attention state: inserting quizzes or interactive content when attention decreases, and pushing more challenging learning tasks when the user is highly focused, maximizing focus utilization. 
In remote collaboration scenarios, real-time feedback on team members' attention states helps optimize meeting schedules and task distribution, ensuring synchronized pacing and boosting overall productivity.

\textbf{Interruption optimization scenario.}
\textbf{AttenTrack II} focuses on identifying users' complete lack of focus, making it suitable for interruption management scenarios that aim for low-cost, high-efficiency information delivery. 
It is particularly applicable in contexts like advertisement delivery, where it minimizes disruptions to users' focused tasks, reducing negative experiences. 
For example, when the system detects a user is unfocused, it can push consumer ads or entertainment content, enhancing interaction conversion rates while minimizing interruption costs.

\textbf{Refined attention detection scenario.}
\textbf{AttenTrack III} enables a three-level granularity of attention state recognition (``low-medium-high''), making it especially suitable for health management platforms that require dynamic adjustment and personalized intervention. 
For example, in mobile health systems, if the system detects that a user has been focused on a single task for an extended period and attention is declining, it can timely send rest reminders or push lightweight content to encourage a shift in focus. 
Additionally, the system can generate personalized reports based on the user’s attention fluctuations throughout the day, helping users optimize their phone usage patterns and establish healthier routines.

\subsection{Limitations and Future Work}
This study uses the ESM method to collect users’ attention state data. 
Although this approach is common in related research, it has certain limitations. 
Specifically, ESM relies on users’ subjective reports, which inevitably introduce individual differences. 
Moreover, the ability to respond to surveys indicates that users are at least somewhat available at the time, making it difficult to capture states of full engagement where they cannot respond~\cite{rosenthal2011using}. 
However, in real-world deployment, AttenTrack treats cases where users are completely unresponsive as highly focused states. 
Our experimental results also demonstrate the effectiveness of cross-user prediction, confirming the practical feasibility of our approach.

Looking ahead, expanding the sensing capability offers further opportunities to improve attention prediction. 
In this study, we rely entirely on smartphones for sensing. However, many users now integrate their smartphones with other devices such as smartwatches and tablets. 
We believe that, in the future, leveraging multi-device collaborative sensing could provide more comprehensive information and enhance the accuracy of predicting users’ attention states.

\section{Conclusion}
\label{sec:conclusion}
Inspired by psychological theories, we attempt to treat mobile notifications as a naturally occurring external distractions, aiming to predict users' attention states by analyzing their responses to these distractions along with contextual information.
Through multiple rounds of field studies, we collect rich user data and explore the relationships between users' attention states, their current context, and responses to external distractions. 
Based on this, we propose AttenTrack, a privacy-friendly, mobile device-based attention awareness model. 
The model can predict the attention state of unknown users without prior user knowledge and demonstrates excellent generalization ability, making it suitable for a wide range of attention management tasks.

\bibliographystyle{unsrt}  
\bibliography{references}

\appendix

\section{Participant Information and Model Details}
\label{sec:appendix_model}

\begin{table}[htbp]
\centering
\caption{Model Parameters for Decision Tree, Random Forest, and Gradient Boosting}
\label{tab:appendix_model_parameters}
\begin{tabular}{|l|l|l|}
\hline
\textbf{Parameter} & \textbf{RandomForest} & \textbf{GradientBoosting} \\ \hline
random\_state & 42 & 42 \\ \hline
class\_weight & balanced & - \\ \hline
criterion & gini & friedman\_mse \\ \hline
max\_depth & None & 3 \\ \hline
max\_features & sqrt & None \\ \hline
max\_leaf\_nodes & None & None \\ \hline
min\_impurity\_decrease  & 0.0 & 0.0 \\ \hline
min\_samples\_leaf & 1 & 1 \\ \hline
min\_samples\_split & 2 & 2 \\ \hline
min\_weight\_fraction\_leaf  & 0.0 & 0.0 \\ \hline
n\_estimators & 100 & 100 \\ \hline
verbose & 0 & 0 \\ \hline
learning\_rate & - & 0.1 \\ \hline
subsample & 1.0 & 1.0 \\ \hline
validation\_fraction & 0.1 & 0.1 \\ \hline
tol & 0.0001 & 0.0001 \\ \hline
\end{tabular}
\end{table}

First, we will briefly introduce the basic information of the participants involved in the two rounds of field research. Table \ref{tab:appendix_participants} displays the participants' basic details.

Our entire field study consisted of two rounds. The first round involves 20 participants aged between 21 and 25, including 19 university students from various majors and one employed individual (holding a Master's degree). Among the students, 16 are graduate students, and three are undergraduate students.
The second round of the field study involves 23 participants, aged between 18 and 55, including eight who had participated in the first round. Among the 23 participants, 18 are students, including 13 graduate students from various majors and five undergraduate students. The remaining five participants are employed in different professions, with educational backgrounds ranging from Bachelor's to Doctoral degrees.
All of our participants use either the Android or HarmonyOS operating systems, with a total of seven different smartphone brands. Our data collection app runs stably on all of these phones.

\begin{table*}[h]
\centering
\caption{The basic information of all participants is as follows. ``r1'' indicates that the participant participated in the first round of the field study. ``r2'' indicates participation in the second round. ``r1, 2'' indicates participation in two rounds. ``Education'' indicates the educational background. If the participant is employed, their highest level of education is recorded; if they are still in school, the current stage of their education is recorded. ``Day'' represents the number of days the participant participated in data collection, and ``Data'' indicates the amount of data provided by the participant.}
\label{tab:appendix_participants}
\begin{tabular}{cccccccc}
\toprule
\textbf{Name} & \textbf{Gender} & \textbf{Age} & \textbf{Education} & \textbf{Study/Work} & \textbf{Phone} & \textbf{Day} & \textbf{Data} \\
\midrule
P1\_r1,2 & Male & 23 & Master's & Studying & Honor & 14, 21 & 122, 521  \\ 
P2\_r1,2 & Female & 22 & Master's & Studying & OnePlus & 14, 21 & 246, 571  \\ 
P3\_r1,2 & Female & 24 & Master's & Studying & OPPO & 16, 20 & 110, 533   \\ 
P4\_r1,2 & Female & 25 & Master's & Studying & HuaWei & 14, 20 & 158, 394   \\ 
P5\_r1,2 & Male  & 23 & Master's & Studying & RealMe & 15, 21 & 191, 280   \\ 
P6\_r1,2 & Male  & 25 & Master's & Studying & Honor & 16, 22 & 175, 602    \\ 
P7\_r1,2 & Male  & 25 & Master's & Studying & OnePlus & 15, 16 & 95, 54  \\ 
P8\_r1,2 & Male  & 23 & Master's & Studying & Honor & 16, 22 & 309, 961  \\ 

P9\_r1 & Female & 25 & Master's & Working & HuaWei & 2 & 16  \\ 
P10\_r1 & Male & 23 & Bachelor's & Studying & IQOO & 10 & 69  \\ 
P11\_r1 & Male & 24 & Master's & Studying & XiaoMi & 3 & 14  \\ 
P12\_r1 & Female & 23 & Master's & Studying & HuaWei & 13 & 100     \\ 
P13\_r1 & Female & 23 & Master's & Studying & OPPO & 3 & 12   \\ 
P14\_r1 & Female & 22 & Bachelor's & Studying & HuaWei & 3 & 34  \\ 
P15\_r1 & Female & 23 & Master's & Studying & HuaWei & 14 & 84   \\ 
P16\_r1 & Male & 21 & Bachelor's & Studying & Redmi & 13 & 199   \\ 
P17\_r1 & Male & 25 & Master's & Studying & HuaWei & 11 & 45  \\ 
P18\_r1 & Female & 23 & Master's & Studying & Redmi & 14 & 94  \\ 
P19\_r1 & Male & 24 & Master's & Studying & Redmi & 15 & 212   \\ 
P20\_r1 & Male & 23 & Master's & Studying & Redmi & 6 & 71   \\ 

P21\_r2 & Female & 23 & Master's & Studying & HuaWei & 12 & 77   \\ 
P22\_r2 & Male & 23 & Master's & Studying & HuaWei & 21 & 323   \\ 
P23\_r2 & Female & 26 & Master's & Working & HuaWei & 8 & 63   \\ 
P24\_r2 & Female & 26 & Bachelor's & Working & Redmi & 19 & 460   \\ 
P25\_r2 & Male & 21 & Bachelor's & Studying & Honor & 20 & 167    \\ 
P26\_r2 & Female & 23 & Master's & Studying & OPPO & 23 & 375    \\ 
P27\_r2 & Male & 22 & Bachelor's & Studying & Redmi & 19 & 247    \\ 
P28\_r2 & Female & 20 & Bachelor's & Studying & IQOO & 7 & 28    \\ 
P29\_r2 & Male & 30 & Doctoral & Working & Redmi & 3 & 9   \\ 
P30\_r2 & Female & 48 & Bachelor's & Working & XiaoMi & 28 & 116   \\ 
P31\_r2 & Male & 24 & Bachelor's & Studying &  Honor  & 10 & 49   \\ 
P32\_r2 & Male & 55 & Bachelor's & Working & HuaWei & 30 & 293   \\ 

P33\_r2 & Male & 21 & Bachelor's & Studying & HuaWei & 19 & 170  \\ 
P34\_r2 & Female & 18 & Bachelor's & Studying & OPPO & 12 & 19    \\ 
P35\_r2 & Female & 23 & Master's & Studying & HuaWei & 22 & 340    \\ 
\bottomrule
\end{tabular}
\end{table*}

We conducted experiments \textit{Random Forest}, and \textit{Gradient Boosting} models from the scikit-learn library, with all models using their default parameter settings without any additional adjustments. The detailed parameter values are listed in Table \ref{tab:appendix_model_parameters}. 

It should be noted that in the testing of the model, some users were excluded because they participated in too few field research days or provided insufficient data (less than 80 data items). Besides, P1\_r2, P5\_r1, P25, P33, and P35 are excluded because their subjective feedback on attention states consisted almost entirely of a single attention level.

\section{GLMM: Random Effects Coefficients of Participants}
\label{sec:appendix_GLMM}
In this section, we primarily discuss the random effects part of the GLMM results. The random effect coefficients are used to capture the variability between individuals or groups, reflecting the degree to which different groups or individuals deviate from the overall model. These coefficients help us understand the variability in response times among different participants.

Table \ref{tab:appendix_GLMM} presents the random effect coefficients for all participants. The coefficients in the table reflect the degree to which participants' response times deviate from the overall trend. By analyzing and comparing these coefficients, we can identify individual differences in response times.

The result shows that there is a significant variation in the random effect coefficients among participants, with the coefficients ranging from -54.54 to 52.18, covering a wide interval. Negative random effect coefficients indicate that the individual's response time is lower than the overall model's average. Conversely, positive coefficients suggest that the individual's response time is higher than the overall model's average.

For example, the random effect coefficient for participant P1\_r1 is -16.13, indicating that his/her response time is significantly lower than the overall model's average. In contrast, participant P19\_r1 has a coefficient of 52.18, indicating that his/her response time is noticeably higher than the overall average. The longer response time could be related to factors such as individual attention allocation or mobile device usage.

In addition, we compared the changes in the random effect coefficients of participants between the first and second rounds of the field study. We found that some participants showed little variation in their random effect coefficients over the four-month period between the two rounds (e.g., P1 and P2). However, there were also participants whose random effect coefficients differed significantly between the two rounds (e.g., P5 and P7). On one hand, this suggests that users' behavioral patterns may change over time; on the other hand, the extent of these changes is largely influenced by individual differences.

\begin{table*}[h]
\centering
\caption{GLMM random effects coefficients for different participants.}
  \label{tab:appendix_GLMM}
\begin{tabular}{|cc|cc|cc|}
\hline
\textbf{Name} & \textbf{Rand. Eff.} & \textbf{Name} & \textbf{Rand. Eff.} & \textbf{Name} & \textbf{Rand. Eff.} \\
\hline
P1\_r1 & -16.1255 & P9\_r1 & -10.9747 & P21\_r2 & -54.5394\\
P2\_r1 & -19.1604  & P10\_r1 & 0.1164 & P22\_r2 & -2.3573\\
P3\_r1 & -5.7652  & P11\_r1 & -25.6771 & P23\_r2 & -21.2375\\
P4\_r1 & -0.0840  & P12\_r1 & 16.5270 & P24\_r2 & 7.4067\\
P5\_r1 & 22.7634  & P13\_r1 & -14.8644 & P25\_r2 & 8.5173\\
P6\_r1 & 2.2976  & P14\_r1 & 12.5454 & P26\_r2 & -2.3084\\
P7\_r1 & 17.0298  & P15\_r1 & 4.5257 & P27\_r2 & 24.8496\\
P8\_r1 & 25.6183  & P16\_r1 & 22.5031 & P28\_r2 & 8.2216\\

P1\_r2 & -10.1522 & P17\_r1 & 18.3344 & P29\_r2 & -21.8730\\
P2\_r2 & -12.8073 & P18\_r1 & -34.2662 & P30\_r2 & 35.1932\\
P3\_r2 & -23.6256 & P19\_r1 & 52.1834  & P31\_r2 & 27.4735\\
P4\_r2 & 32.7212 & P20\_r1 & -37.2641  & P32\_r2 & 16.6249\\
P5\_r2 & -13.1996 & &  & P33\_r2 & -26.2671 \\
P6\_r2 & 35.6025 &  &  & P34\_r2 & -27.0041\\
P7\_r2 & -30.9991 & & & P35\_r2 & -14.5281 \\
P8\_r2 & 34.0251 & & & & \\
\hline
\end{tabular}
\end{table*}

\end{document}